\documentclass[usenatbib]{mn2e}

\usepackage{amsmath}
\usepackage{ulem}
\usepackage{amssymb}
\usepackage{float}
\usepackage{graphicx}
\usepackage{lscape}
\usepackage{url}
\usepackage{natbib}

\title[MHD Shear Flows]{Molecular cloud formation in high-shear, magnetized colliding flows}

\author[E. Fogerty et al.]{E.  Fogerty$^{1}$\thanks{E-mail:erica@pas.rochester.edu}, A. Frank$^{1}$, F. Heitsch$^{2}$, J. Carroll-Nellenback$^{1}$, C. Haig$^{2}$, M. Adams$^{1}$ \\\\ 
$^{1}$
206 Bausch \& Lomb Hall, Department of Physics \& Astronomy, University of Rochester, Rochester, New York, 14627, United States\\
$^{2}$
3255 Phillips Hall, Department of Physics \& Astronomy, University of North Carolina, Chapel Hill, North Carolina, 27599, United States}

\begin{document}

\date{Submitted 2015 February 28}

\pagerange{\pageref{firstpage}--\pageref{lastpage}} \pubyear{2015}

\maketitle

\label{firstpage}

\begin{abstract}
The colliding flows (CF) model is a well-supported mechanism for generating molecular clouds. However, to-date most CF simulations have focused on the formation of clouds in the normal-shock layer between head-on colliding flows. We performed simulations of magnetized colliding flows that instead meet at an oblique-shock layer. Oblique shocks generate shear in the post-shock environment, and this shear creates inhospitable environments for star formation.  As the degree of shear increases (i.e. the obliquity of the shock increases), we find that it takes longer for sink particles to form, they form in lower numbers, and they tend to be less massive. With regard to magnetic fields, we find that even a weak field stalls gravitational collapse within forming clouds. Additionally, an initially oblique collision interface tends to reorient over time in the presence of a magnetic field, so that it becomes normal to the oncoming flows. This was demonstrated by our most oblique shock interface, which became fully normal by the end of the simulation.


\end{abstract}


\section{Introduction}

The interstellar medium (ISM) is a dynamic environment where material cycles from a warm, tenuous phase to a cold, dense phase and back again. The evolution of the gas through these phases establishes limits on star formation in the galaxy. Understanding the details of such dynamics is crucial as theoretical scenarios for molecular cloud evolution (and hence star formation) have shifted away from models based on long molecular cloud lifetimes (i.e. steady states where $\tau_{cloud} >> \tau_{ff}$; cf. \cite{shu1987, mouschovias1991}) .
Over the last decade or more, evidence has grown that instead supports a scenario in which clouds are {\it transient} structures born out of the cold, dense phase of the ISM. For example, starless clouds appear to be scarce, meaning clouds do not slowly evolve towards conditions in which star formation begins \citep{beichman1986, dame2001}.  Also, the majority of stars in clouds are, in general, young showing ages $< 5 ~Myr$ \citep{hartmannetal2001}. 
This implies that star formation begins soon after the cloud itself forms and that clouds are not generally long-lived due to the absence of older stars \citep{fukui1998, palla2000, carpenter2000}. Finally, high degrees of hierarchical structure in molecular clouds should have short lifetimes due to star-star scattering or tidal interactions \citep{lada1995, eisenhauer1998, beck1998}. 

While observational support for short cloud lifetimes has grown steadily, we note that scenarios invoking rapid cloud and star formation are not new \citep{hunter1979, larson1981, hunter1986, ballesterosparedesetal1999}. Given the complexity that comes with a dynamical theory of molecular cloud evolution, however, exploring the full dynamics of these scenarios depends heavily on high performance computational methods. One scenario that generates transient molecular clouds and whose exploration has been possible with modern numerical simulations is the 'colliding flows' model of molecular cloud formation. In this model, molecular clouds are formed in the shocked collision layer between two large-scale colliding streams of gas. Simulations of colliding flows have shown that nonlinear density structures readily form in the shocked collision region between the flows via a variety of instabilities (cf. \cite{heitsch2005} for discussion of the unstable modes). Further, these structures develop column densities high enough for effective UV-shielding, and thus, $H_2$ formation ($N\approx 1-2 ~cm^{-2}$; \cite{vandishoeck1988, vandishoeck1998}). Gravitational instabilities then cause the turbulent, shocked gas in the dense structures to collapse and form stars. Given the highly dynamical environment, the transition from the beginning of molecular cloud formation to star formation to cloud destruction occurs in roughly a dynamical time  \citep{audit2005, heitsch2006, vazquez-semadeni2006, heitsch2008}, matching observations (\cite{elmegreen2000, ballesteros-paredes2007}, and references therein). The clouds produced in colliding flows simulations are similar in many ways to those seen in observation.

While colliding flows models tend to be idealized, they are not without motivation. Coherent large-scale streams of gas are plausible in many situations in the ISM, such as, expanding bubbles driven from energetic OB associations and/or supernovae, turbulent motions arising from gravitational instabilities, density waves in the spiral arms of galaxies, and cloud-cloud collisions \citep{hartmann2001b, inutsuka2015}. Observational evidence supporting these various scenarios, as well as their association with molecular cloud formation, takes a number of forms. Atomic inflows surrounding molecular gas have been observed in Taurus \citep{ballesteros-paredes1999} and other molecular clouds \citep{brunt2003}. \cite{looney2006} show that the active star forming molecular cloud core in BD +40 4124 likely arose from cloud-cloud collisions (a localized version of a colliding flow). Molecular clouds have been found at the edges of supershells \citep{dawson2011, dawson2013}, which are a form of colliding flows driven by multiple supernovae.

If molecular clouds form via the accumulation of gas on dynamical timescales \citep{hartmannetal2001}, then magnetic fields are expected to be dynamically important on similar timescales. Some colliding flows simulations have studied the role of magnetic fields in molecular cloud formation \citep{heitsch2007, hennebelle2008, banerjee2009, heitsch2009, vazquez2011, chen2014, kortgen2015}. Some of these simulations \citep{heitsch2007, hennebelle2008, banerjee2009} have found power law relations between the magnetic field strength and density, similar to what is seen observationally, i.e. $B \propto n^k$, where $1/2<k<2/3$ for $n>100 ~cm^{-3}$, and $k=0$ for $n<100 ~cm^{-3}$ \citep{troland1986, crutcher1999, crutcher2010, tritsis2015}. The inclusion of fields in the models also reduces the star formation rates to values more in agreement with observations \citep{vazquez2011, chen2014}. Such a reduction has been claimed to result from lower degrees of turbulent substructure found in the simulations \citep{heitsch2007, hennebelle2008, heitsch2009, chen2014}.  

While the colliding flows model has proved useful for understanding transient molecular cloud formation, the great majority of work has focused on head-on collisions, with no obliquity in the initial shocks formed in the flow. It is therefore worthwhile to explore the consequences of allowing the flows to interact at an interface initially inclined relative to incoming velocity vectors.  Strong shear in the interaction region can lead to stretching of embedded field lines and the possible generation of turbulence from KH modes. The role of shear in molecular cloud and star formation has begun to be investigated numerically. Hydrodynamic simulations by \cite{rey-raposo2015} show that clouds can inherit shear velocity fields during their formation in spiral arm galaxies, and that these shear flows impair subsequent star formation. \cite{kortgen2015} find that varying the intersection angle of magnetized colliding flows reduces the star formation efficiency of the gas. This was attributed to a post-shock shear flow disrupting the formation of high density structures. 

Continuing along these lines, we address the role of shear and magnetic fields in molecular cloud formation, with a focus on the bulk dynamics of the flow. Shear is generated in our models by keeping the flows parallel and varying the angle of the collision interface. We present four adaptive mesh simulations, at a peak resolution of $0.05~pc$, varying the collision interface from a normal incidence to highly inclined. All of our simulations include a uniform magnetic field aligned with the flows to simulate idealized ISM conditions.  We also include self gravity and time-dependent cooling. We present our numerical model in Section \ref{model}, our results in Sections \ref{m2fr}-\ref{spectra}, and our discussion in Section \ref{discussion}.

\section[]{Numerical Model}\label{model}

\begin{figure}
\includegraphics[width=\columnwidth]{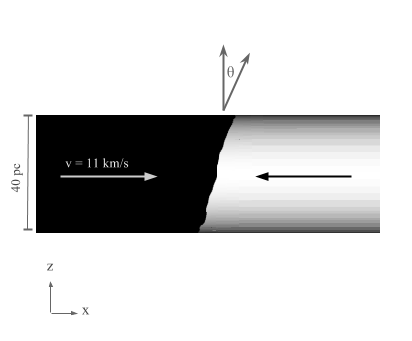}
\caption{Model diagram. Two oppositely driven cylindrical flows intersected at an angle $\theta$, rotated about the $y-$ axis. The interface was perturbed with a random sequence of sine waves. The flows were homogeneous and embedded in a stationary, uniform ambient medium of the same density and pressure. The setup was initialized with a uniform magnetic field in $x$, parallel to the direction of the colliding flows.}
\label{fig1}
\end{figure}

We conducted our simululations using  AstroBEAR\footnote{https://astrobear.pas.rochester.edu/trac/} \citep{carroll2013}, a publicly available, massively parallelized, adaptive mesh refinement (AMR) code that contains a variety of multiphysics solvers (i.e. self-gravity, magnetic resistivity, radiative transport, ionization dynamics, heat conduction, and more). Our setup was of two, $40~pc$ diameter cylinders colliding in a 3D domain under the influence of gravity, magnetic fields, and cooling (Fig. \ref{fig1}). Gravity and cooling source terms were solved using a Strang-split corner upwind transport (CTU) scheme that was 2nd-order accurate in time. This was combined with a directionally unsplit CTU scheme for the 3D ideal magnetohydrodynamics (MHD) equations, using the HLLD Riemann solver. Gravitational interactions included both the self-gravity of the gas, as well as the gravitational acceleration due to sink particles, which were implemented following \cite{fedderrath2010}. To solve Poisson's equation for the gravitational potential of the grid, AstroBEAR uses {\sevensize{HYPRE}}\footnote{HYPRE is a software package that solves linear systems on massively parallelized systems. Documentation on Hypre can be found at: https://computation.llnl.gov/casc/hypre/software.html} 
(\cite{falgout2002}; see the appendix in \cite{kaminski2014} for a description of AstroBEAR's self-gravity algorithm).

A uniform magnetic field was initialized everywhere in the box, parallel to the flows. The field was initially dynamically weak, with $\beta=10$ and  $\beta_{ram}\approx 38$ at the start of the simulations ($\beta$ is the ratio of thermal to magnetic pressure, and $\beta_{ram}$ is the ratio of ram to magnetic pressure). The field had an initial strength of $B = 1.3 ~\mu G$, which is at the lower end of current ISM magnetic field estimates \citep{beck2001, heiles2005}. Cooling and heating  were included using a parametrized cooling curve adapted from \cite{inoue2008} to include the effects of UV shielding. The modification allowed the gas to cool to $T=10 ~K$ for densities greater than $n > 1000 ~cm^{-3}$ (Ryan \& Heitsch in prep). 

Our simulations tracked the formation of gravitationally collapsed objects (sink particles) which were created only after a series of checks had been satisfied, e.g., 

\begin{enumerate}
   \item a cell and surrounding region was Jeans unstable
   \item the center cell (i.e. location of potential sink) in this Jeans unstable region was collocated with a gravitational potential minimum
   \item the surrounding region was exhibiting a converging flow into this center zone
\end{enumerate}
etc., as given in \cite{fedderrath2010}. These checks ensured that sinks formed only when appropriate, i.e. when a region would go on to form a gravitationally bound object if better resolution were available. The simulations had 5 levels of AMR, giving a finest cell size of $\Delta x_{min}=0.05 ~pc$. Given this resolution, we identified single sink particles as proto{\it clusters} rather than protostars. We will therefore use the words 'sink' and 'protocluster' interchangeably. 

Sink particles interacted with the gas (and other sinks around them) through gravitational interactions, and the ability to accrete surrounding material. They remained individual objects throughout the course of the simulation (i.e. did not merge), and did not provide any form of energy or momentum feedback into the surrounding medium (i.e no winds or radiation). Gas around a sink particle was accreted only when the density in the surrounding zones exceeded a given threshold, dictated by the Truelove condition \citep{truelove1997}, 

\begin{equation}
  \lambda_J > 4 \Delta x_{min}
\end{equation}

\noindent where $\lambda_J$ is the cell-centered Jeans length. Then, only the excess gas was removed \citep{fedderrath2010}. 

The suite of simulations consisted of four runs, each with a different orientation of the collision interface, specified by the inclination angle $\theta$ (Table \ref{tab1}). The inclination angle was varied between $\theta=0 \degr$ and $\theta=60\degr$, i.e. between a head-on and highly inclined collision. This translated into the difference between a normal shock at the collision layer and an oblique shock. All parameters for the present suite of runs were the same as the 'smooth' model of \cite{carroll2014}, with the exception of the magnetic field and the variation of the collision interface (the smooth run was a hydro, head-on collision). We will therefore compare the $\theta=0\degr$ run to the smooth run to study the effect of the magnetic field alone. Hence, the smooth run from here on out will be called the 'hydro version of the $\theta=0\degr$ case' (or, the 'hydro run', for short).

\begin{table}
\centering
\caption{The suite of simulations. $\theta$, $L_x$, $L_y$, $L_z$, and $t_{sim}$ denote the inclination angle, box dimensions, and final simulation time, respectively. The box size was increased in two of the runs to accommodate the steeper angle. The final simulation time was extended in the $\theta=60\degr$ case to check for sink particle formation.}
\label{tab1}
\begin{tabular}{@{}clclclc@{}}
\hline
 $\theta$ ($\degr$)  & &  $L_x ~(pc)$  & & $L_y, ~L_z ~(pc)$ & & $t_{sim} (Myr)$ \\
\hline
 0 & & 62.5 & & 75 && 27.3   \\
 15  & & 62.5 & & 75 &&  27.3 \\
 30 & & 200 & & 75 && 27.3 \\
 60 & & 200 & & 75 && 32.8 \\
\hline
\end{tabular}
\end{table}

The flows were injected into a stationary ambient medium at a velocity of $v=11 ~km ~s^{-1}$ and an isothermal mach of $M=1.5$. The mean molecular weight was set to $\mu=1.27$, and the adiabatic exponent to $\gamma=5/3$. The gas was initially in thermal equilibrium at a uniform number density of $n=1~cm^{-3}$, corresponding to the (linearly) stable, warm neutral medium (WNM) phase of the cooling curve. This set the thermal pressure and temperature everywhere inside the box to be $P_{therm} ~k_B^{-1}=4931 ~K~cm^{-3}$, and $T=4931 ~K$, where $k_B$ is the Boltzmann constant. The ram pressure of each flow was $P_{ram} k_B^{-1}=18,500 ~K~cm^{-3}$, giving a total mass flux into the collision region of $M_{flux}\approx 886~ M_{\sun}Myr^{-1}$. 

As in previous simulations, the collision interface was seeded with a set of random sinusoidal perturbations to excite the nonlinear thin shell (NTS) and Kelvin Helmholtz (KH) instabilities \citep{heitsch2006, carroll2014}. These perturbations had a maximum amplitude of $A=2 ~pc$, spectral index $\alpha= -2.0$, and maximum wave number $k_{max}=16~pc^{-1}$. Boundary conditions on the box were set to inflow-only on the faces where the flows were injected, and extrapolating on all other faces. The boundary conditions for the gravity solver were set to multipole expansion. 
 
The simulations were initialized to 3 levels of AMR, with the finest cells centered on the collision interface within a cylindrical volume $40 ~pc$ in diameter and $20 ~pc$ long, and the coarser meshes nested outwards from there. This made for an {\it initial} effective resolution of $\Delta x_{eff} = 0.2 ~pc$. Two additional levels were triggered throughout the simulation based on gradients in the fluid variables, as well as resolution of the local Jeans length ($\lambda_J$), such that if a cell's $\lambda_J$ was smaller than 64 zones on that cell's level, another level of AMR would be added. This brought the finest cell size to $\Delta x_{min} = 0.05 ~pc$, as previously stated. The final simulation time for all of the runs was $t_{sim} = 27.3 ~Myr$, with the exception of the $\theta = 60 \degr$. This was the only run that did not form any sink particles by this time, and so was extended out until $t_{sim} = 32.8 ~Myr$. 
 
In what follows, our analysis will focus on the $\theta=0$, $15$, and $60\degr$ cases, as the $\theta=30\degr$ case did not significantly differ from the $\theta=15\degr$ case.

\section{Generating shear via an oblique collision interface}

The jump conditions across an oblique, 1D, shock-bounded slab convert initially intersecting velocity vectors into a shear flow field. As a measure of this shear for our setup here, we produced mass weighted histograms of the magnitude of the vorticity ($||\mathbf{\nabla}\times\vec{v}||$) in cylindrical analysis regions centered on the collision region. Each of these "hockey pucks" encompassed the corresponding collision region by tracing the collision interface and extending out to $5 ~pc$ on either side of the interface. They were normalized to contain the same amount of mass ($1,000 M_{\sun}$). Histograms were generated at $t=1 ~Myr$, which is approximately the time it would take a post-shock sound wave to travel from the center of the CF cylinder to the outer boundary. 

\begin{figure}
\includegraphics[width=\columnwidth]{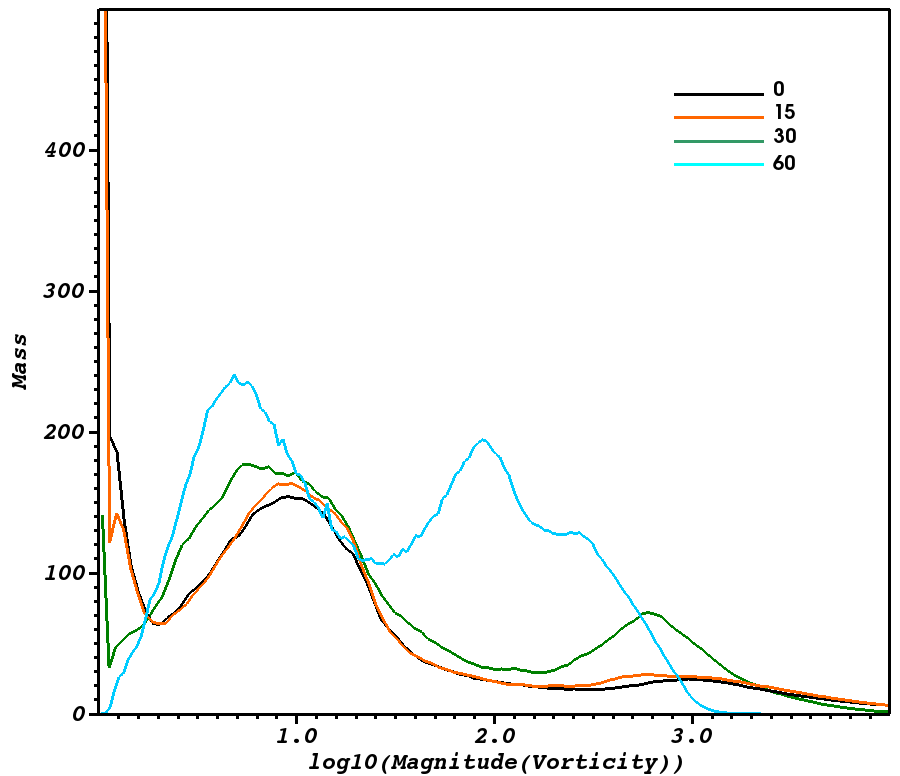}
\caption{Mass-weighted vorticity histograms at $t=1~Myr$. The legend gives the inclination angle ($\degr$) of the given run. Each of the histograms were binned from a cylindrical analysis region centered on the given collision interface (see text), and were normalized to contain $1,000 ~M_{\sun}$. Units are scale-free, where the scale factors are given by $MM^{-1}_{\sun}=100$, $ll^{-1}(pc)=1$, $v v^{-1}(cm s^{-1})=8091$.}  
\label{vort}
\end{figure}

As can be seen in Fig. \ref{vort}, lower shear runs (i.e., $\theta=0-15\degr$) have significantly more mass at lower vorticity ($||\mathbf{\nabla}\times\vec{v}||<100$), than at higher. As the collision angle increases to $\theta=30 \degr$, we see two changes occur. First, the amount of mass at $||\mathbf{\nabla}\times\vec{v}||\leq1$ decreases considerably. Second, more of the mass moves to higher vorticity. As the collision angle steepens to $60\degr$, this trend strengthens. For the $\theta=60\degr$ case, there is neglibible mass at the lowest $||\mathbf{\nabla}\times\vec{v}||$. That is to say, nearly all of the mass has acquired vorticity in this run. Moreover, \textit{most} of the mass has acquired high vorticity ($||\mathbf{\nabla}\times\vec{v}||>10$). 

Thus, as the inclination angle increases, more vorticity is generated in the collision layer. This vorticity can be associated with the solenoidal mode of the post-shock turbulence \citep{federrath2010b}. Turbulent solenoidal fluid motions are efficient in providing support against collapse. Compared to compressive modes of turbulence, \cite{federrath2012} show that solenoidal modes greatly reduce the star formation rate in turbulent, magnetized clouds. The passage of gas through the oblique shocks of our simulations thus transforms the compressive nature of the flows into solenoidal. In this way, our higher shear cases should exhibit greater degrees of turbulent support.

\section{Mass-to-flux ratio}\label{m2fr}

Before we present the results of the simulations, we will first discuss some general issues associated with gravitational stability and magnetic fields that are relevant to our studies. We begin with the critical mass-to-flux ratio (M2FR) of cylindrical uniform flows. The M2FR compares the relative strength of the magnetic field to gravity. It does not take into account other forces which oppose gravity, such as thermal or ram pressure forces. The M2FR is given by,

\begin{equation}\label{m2fr_eqn}
  \mu_{crit}=\frac{\Sigma}{B}\approx \frac{1}{\sqrt{4 \pi^2 G}}
\end{equation}

\noindent where $\Sigma$ is the mass column density and $B$ is the magnetic field threading the cylinder \citep{nakano1978}. Equation \ref{m2fr_eqn} can be rearranged for the {\it critical length} of the cylinder. In terms of typical ISM values this is given by,

\begin{equation}\label{lcrit_eqn}
  L_{crit}\approx 470~pc  ~(\frac{B}{5 \mu G})~(\frac{n}{cm^{-3}})^{-1} 
\end{equation}

\noindent where $n$ is the number density \citep{vazquez2011}. For the initial WNM values of 
our flows ($n=1 ~cm^{-3}$, $B=1.3~\mu G$), Eqn. \ref{lcrit_eqn} gives a critical length of,

\begin{equation}\label{lcritWNM_eqn}
 L_{crit, WNM}\approx 122 ~pc
\end{equation}

\noindent for the WNM component of the gas. Using this, we can ask whether we would expect the warm gas to be magnetically supercritical. We first have to define a length scale over which we will check stability. Naturally this would be the collision region, as this is where the molecular clouds will be forming. As we will see, the width of this region ($L_{coll}$) is on the order of $10 ~pc$. This means that over the length scale of the collision region, the flow would be {\it sub}-critical. That is, given $L_{crit} >> L_{coll}$, the magnetic field, at least initially, should be strong enough to withstand gravity.

Given the mass flux into the collision region along the colliding flows, one can ask '{\it how long} will it take to accumulate enough mass into the collision region for the collision region to go unstable?' For this, it is helpful to recast Eqn. (\ref{lcrit_eqn}) in terms of the critical column number density,

\begin{equation}\label{ncrit_eqn}
  N_{crit}\approx 1.45 \times 10^{21} (\frac{B}{5 \mu G}) ~cm^{-2}
\end{equation}

\noindent \citep{vazquez2011}. For simplicity, we imagine the collision region to be initially massless. Its column density as a function of time is then,

\begin{equation}\label{col_dens}
 N(t) = 2 n v t
\end{equation}

\noindent where $v$ is the speed of the flows. Equating this to Eqn. (\ref{ncrit_eqn}) and solving for $t$ gives the timescale for the collision region to become magnetically supercritical: $t_{crit}\approx 6~Myr$. Thus, the collision region is expected to very quickly (i.e. $t_{crit} << t_{sim}$) become magnetically supercritical. While  this is a crude estimate, when taken in an average sense (over the collision region) it implies that the {\it mean} field is unable to support the collision region against collapse. However, despite this prediction of a weak field, we did not see a large-scale, global collapse occur in our simulations. This, we will argue, was due to the kinetic energy of the flows themselves, which also prevented global collapse in the hydro version of the $\theta=0\degr$ case \citep{carroll2014}, rather than the magnetic field. Thus, the M2FR should be used with caution for estimating global stability within the colliding flows model.

We next consider smaller scales associated with the cold, dense phase out of which molecular clouds form. The cold neutral medium (CNM) in our simulations had number densities of approximately $n\approx 500 ~cm^{-3}$. For the same initial uniform magnetic field, the critical length for the CNM is,

\begin{equation}\label{lcritcnm_eqn}
  L_{crit,CNM }\approx 0.2 ~pc
\end{equation}

\noindent As we will see, column density structures associated with the CNM were typically much larger in width than this. Thus, widespread {\it local} collapse (i.e. over length scales associated with the cold component of the gas) might be expected to occur in our simulations.

However, we did not see widespread local collapse. Clearly, the magnetic field assumption in the M2FR calculation was in error. Turbulent instability behind the shocks would have deformed the magnetic field, thereby producing strong field fluctuations. We suspect that field amplification within the cold, dense gas inhibited collapse {\it locally}, and that only after the excess magnetic energy was lost (e.g. through numerical reconnection), could collapse proceed. Thus, we expect the cold clumps were actually largely {\it sub}-critical and that this could have been shown through a more rigorous calculation of the local M2FR. While such a calculation was beyond the scope of this paper, others have worked on calculating the local M2FR in simulations (e.g. \cite{banerjee2009,vazquez2011, chen2014}).

Thus, in what follows we will be concerned with issues of local vs. global collapse and the mechanisms by which other processes, such as, turbulence and field amplification, can inhibit collapse. In particular, we must consider that for large values of the inclination angle $\theta$, post-shock flows can retain a significant fraction of their pre-shock velocity.  As shear leads to turbulence, we expect the turbulent velocity to be some fraction of the incoming velocity ($v_{turb}\propto f v_o$).  If turbulence provides support to the cloud against collapse, then we would expect enhancements of the local Jeans length due to turbulence ($\lambda_{turb}$) to be of order,

\begin{equation}\label{}
\frac{\lambda_{turb}}{\lambda_{J}} \propto \frac{fv_o}{c_{s}(\bold{x})}
\end{equation}

\noindent where $c_s(\bold{x})$ is the local sound speed. Thus, depending on the fraction of inflow velocity retained by the turbulence, we expect the collision region in flows with shear to be more stable to collapse than those without shear.  In addition, any degree of turbulence will produce local field  amplifications (i.e. additional support against collapse), which as stated, was not accounted for by the M2FR analysis presented above.

\section{Protocluster formation and evolution}\label{massform}

In this section we summarize our principle result by demonstrating how shear and magnetic fields directly affected the star formation in terms of the creation of sink particles. While the sink particles in our simulations had masses that should be associated with clumps (i.e. protoclusters), were we to include higher levels of resolution we would expect the clumps to form cores which would then collapse to form individual stars.

\begin{figure}
\includegraphics[width=\columnwidth]{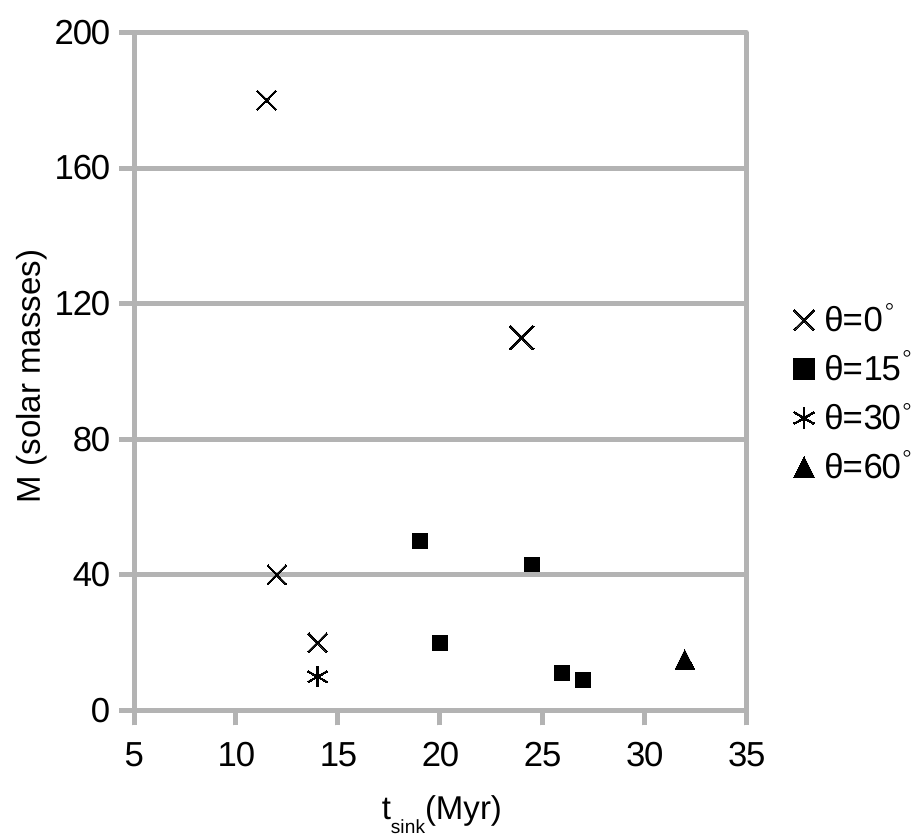}
\caption{Final mass distribution of protoclusters as a function of formation time, $t_{sink}$. Note, the final simulation time was the same for the $\theta=0, 15,$ and $30\degr$ runs ($t_{sim}=27.3 ~Myr$), but was longer for the $\theta=60\degr$ case ($t_{sim}=32.8$). Also, mass accretion was not constant over time, but rather depended on environmental conditions.} 
\label{fig2}
\end{figure}

Fig. \ref{fig2} presents the final masses of the sink particles as a function of their formation time for the various runs. This plot relays three pieces of information that show the effect of shear. First, the total number of protoclusters formed in each of the runs {\it decreased} with shear. The number decreased from four protoclusers in the $\theta=0\degr$ case to one in the $\theta=30\degr$ case to zero in the $\theta=60 \degr$ case (in the run time given to the other simulations). Only after extending the $\theta=60\degr$ case out another $5~ Myr$ did a protocluster eventually form at $t_{sink}(\theta=60\degr) \approx 32 ~Myr$, as shown  in Fig. 2. While this trend did not hold for the  $\theta=15\degr$ case, which formed a couple of low mass protoclusters late in the simulation, it is clear that the initial inclination angle of the colliding flows interface was directly related to the amount of post-shock support against collapse. 

Second, higher levels of shear {\it delayed} the formation of protoclusters. The $\theta=0\degr$ case produced a protocluster by $t\approx 11~ Myr$, whereas the $\theta=15 \degr$ case did not produce a protocluster until $t\approx 18 ~Myr$. As mentioned above, the ($\theta=60 \degr$)  was inhospitable enough that protocluster formation was delayed until $t \approx 32 ~Myr$. This shows the various simulations were evolving under different timescales. This can be understood by considering $t_{crit}$ - the timescale to acquire enough material into the central region to go magnetically supercritical. As the inclination angle was varied, the velocity in Eqn. \ref{col_dens} would have become $v' = v\cos{ \theta}$, due to deflection at the collision interface (i.e. the generation of shear). Thus, the critical timescale for shear environments is related to $t_{crit}$ by,

\begin{equation}
  t_{crit}' \propto t_{crit}(\cos{\theta})^{-1}
\end{equation}

\noindent This equation predicts longer timescales for gravitational instability in higher shear environments, consistent with the increased time to form protoclusters and the overall reduction in protocluster number. Finally we note that while the $\theta=30 \degr$ case does disagree with this result (forming its single protocluster before the $\theta=15\degr$ case), this sink particle was extremely low-mass despite having ample time to grow. This indicates that while local collapse occurred in the region where the protocluster formed, the shearing motions generated by the initial inclination angle of the interface inhibited further growth. 

Third, shear {\it slowed} growth of protoclusters, i.e. higher shear cases had lower average accretion rates ($<{\dot{M}}>=M_{final}/(t_{sim}-t_{sink})$). For example, consider the two protoclusters that formed at $t\approx 24 ~Myr$. For this pair, the $\theta=0\degr$ protocluster grew to be $\approx 3\times$ more massive than its $\theta=15\degr$ counterpart. This indicates that the environments surrounding the protoclusters were less bound in the higher shear runs (recall, only gravitationally bound and unstable gas in the surrounding zones can be accreted onto sink particles). The higher accretion rates of the lower shear simulations translated into more massive protoclusters. 

The role of magnetic fields on dynamics can also be extracted from Fig. \ref{fig2}, as there was a significant reduction in the number of protoclusters formed in the MHD, no-shear case ($\theta=0\degr$), compared to the hydro version of this run \citep{carroll2014}. For the same final simulation time and resolution, the hydro run formed a total of 27 protoclusters compared to the 4 formed in the $\theta=0\degr$ case. Moreover, \cite{carroll2014} found differences in the mass distributions of protoclusters depending on whether there was global or local collapse. The present protocluster masses are consistent with the mass distribution of the hydro run, which only exhibited local collapse. Taken together, this demonstrates the local support magnetic fields are providing against collapse, since neither the hydro or MHD runs showed evidence of global collapse. 

\begin{figure}
\includegraphics[width=\columnwidth]{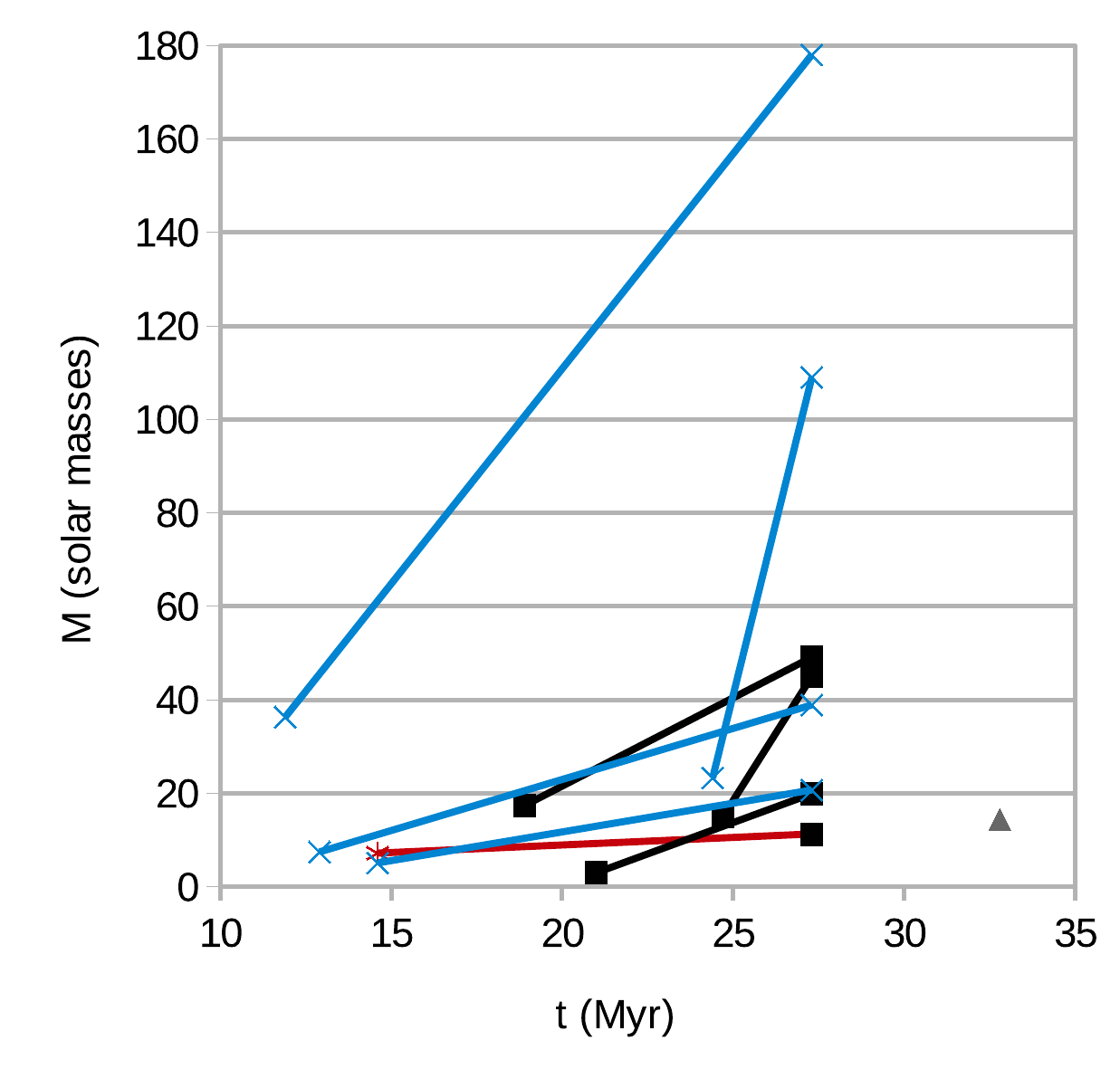}
\caption{Masses of the various sink particles  at $1~Myr$ after their formation with corresponding final masses. Each set of points (early and final mass), is connected by a line. The $\theta=0,15,30,60\degr$ cases are given by the blue, black, red, and grey lines, respectively. Note that the last sink particle for the $\theta=15\degr$ case is not plotted, as that sink's lifetime was less than $1~Myr$. The $60\degr$ is a single point as that sink lived for exactly $1~Myr$.}
\label{massat1myr}
\end{figure}

\begin{figure}
\includegraphics[width=\columnwidth]{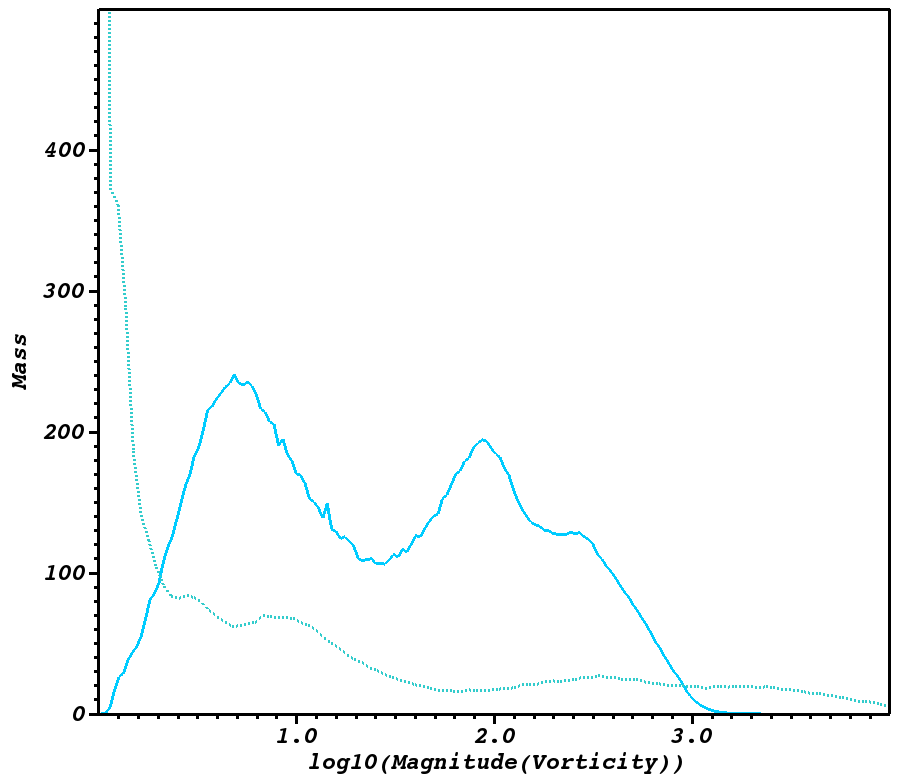}
\caption{Mass-weighted vorticity histograms for the $\theta=60 \degr$ case at early and late times. Solid line is at $t=~1 ~Myr$, dotted is at the end of the simulation ($t=32.8~Myr$). Each of the histograms were binned from a cylindrical analysis region that was tilted $60\degr$ and contained $1,000 ~M_{\sun}$. Units are again scale-free.}
\label{s60vort}
\end{figure}

We now turn to comparing the sink particles at a similar evolutionary time. Figure \ref{massat1myr} shows the mass distribution of the sink particles at $1 ~Myr$ post-formation. This time was chosen as it maximizes the accretion time of the $\theta=60\degr$ sink. However, this time loses the final sink of the $\theta=15\degr$ run, which had a lifetime $< 1 ~Myr$. In the figure, each sink's mass at $1~Myr$ post-formation is connected by a line to the final mass of that sink. This provides a quick reference of the average accretion rates of the sink particles. Figure \ref{massat1myr} shows that both the initial peak mass, as well as the average initial mass, decline with increasing shear (focusing on the left-most points of each pair). It is interesting that the $\theta=60\degr$ case seems to contradict this behavior. However, this has to do with the reorientation of the collision interface (discussed in more detail below). By the time this sink particle forms, the initially steep collision angle had become largely normal to the oncoming flows. This would have reduced the degree of post-shock shear, and consequently the amount of support afforded to the surrounding envelope. Thus, the $\theta=60\degr$ sink particle was able to accrete at a rate comparable to lower shear runs. Mass-weighted histograms of the vorticity in the collision region for the $\theta=60\degr$ case shows that this has indeed occurred (Fig. \ref{s60vort}).

To conclude this section, we note that the accretion rates of the various sink particles were highly environmentally dependant. Some sinks were positioned in large regions of gravitationally unstable, collapsing gas compared to others, and this would have caused their accretion rates to be larger. This is why some of the sinks that have formed later in time have grown to be more massive than older sinks.

\section{Morphology}\label{morphology}

We now turn to column density maps (CDMs) for a more detailed comparison of the flow evolution. We begin by discussing a morphological artifact common to MHD colliding flows, and then move on to the main morphological features of the flow. 

\begin{figure*}       
\includegraphics[width=\textwidth]{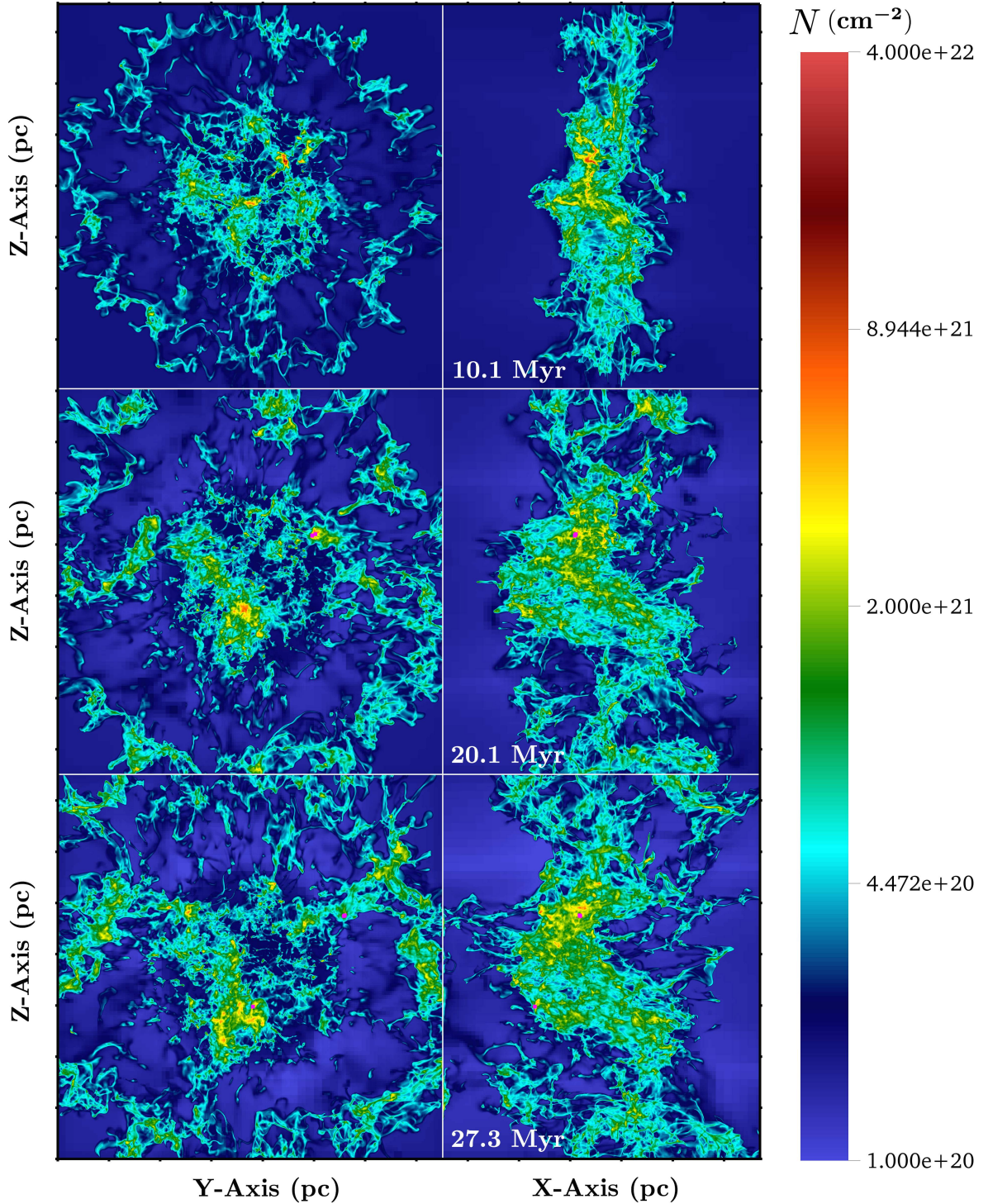} 
\caption{Column density map for the $\theta=0\degr$ case. The integration length is the $62.5 ~pc$ for the left-hand column and $75 ~pc$ for the right-hand column in these and all subsequent column density maps. Sink particles are given as fuchsia points. Each tick mark represents $10 ~pc$.}
\label{fig3}
\end{figure*}

Column density maps of the $\theta =0 \degr$ case are shown in Fig. \ref{fig3}. As can be seen in the left hand panel, there is a large, ring-like structure surrounding the flows. This ring is an artifact of the simplified initial conditions and its formation can be understood as follows. Initializing colliding flows as cylinders naturally produces a region where shocked gas is expelled laterally with respect to the cylindrical axis. Any colliding flows geometry will produce a characteristic two shock structure (one to decelerate each flow), separated by a contact discontinuity.  For finite-sized flows, there must also be a region where high pressure post-shock material is driven out of the collision region. Analysis of a similar process in time variable protostellar jets (which, with a change in reference frame are similar to the configuration studied here), shows that lateral motions on the order of the post-shock sound speed ($c_{ps}$) carry material away from the interaction region along radial streamlines into the ambient gas \citep{raga1992}. In this way, converging flows along the length of the cylindrical regions are converted into radial flows expanding away from the axis of the cylinders.  

\begin{figure}    
\includegraphics[width=0.8\columnwidth]{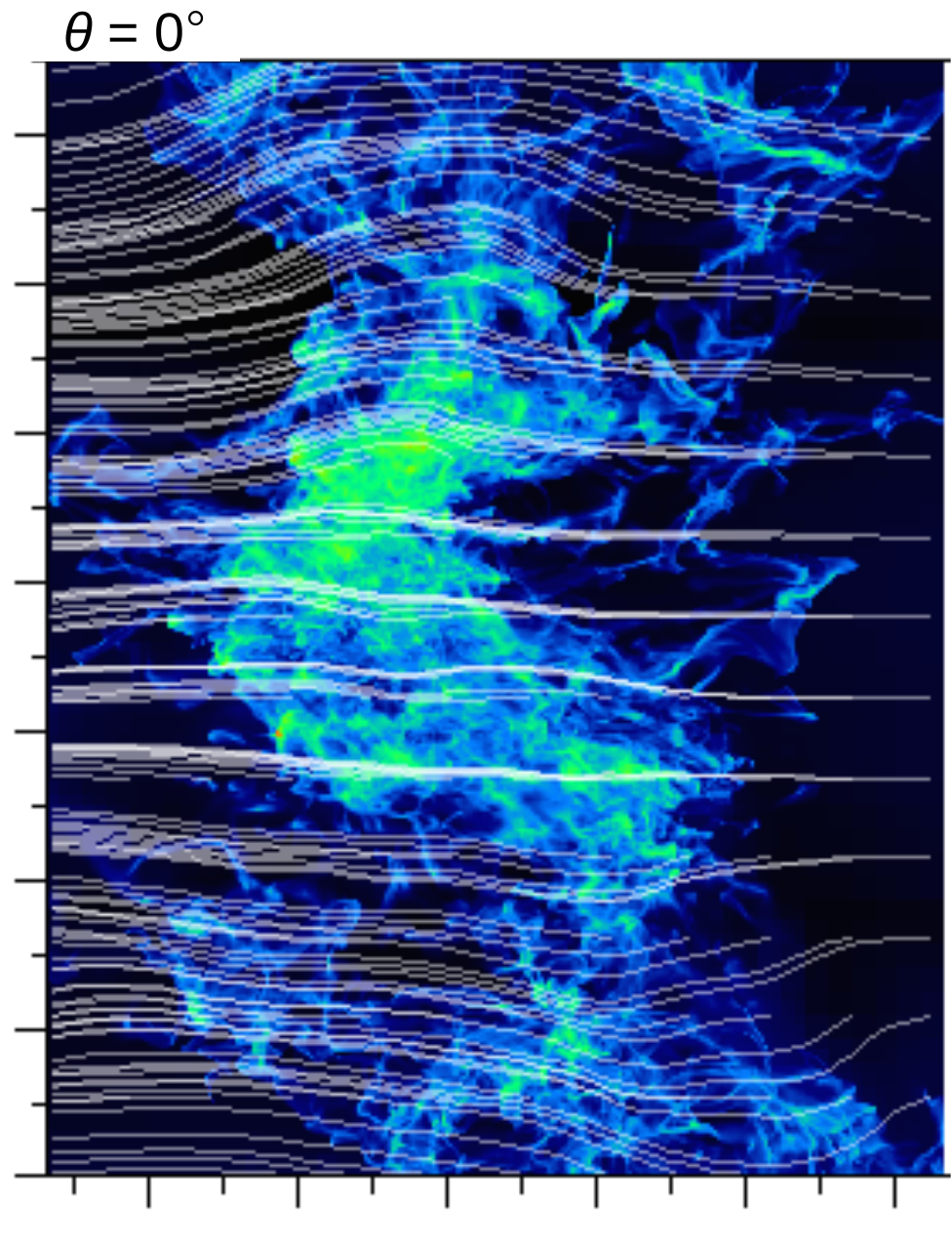}
\caption{Streamline plot of $B_x$ and $B_z$ components of the magnetic field averaged along $y$ as shown in the $x-z$ plane for the $\theta=0\degr$ case. Note the bending of the field lines as material is expelled from the collision region (see text for description).}
\label{s0streams}
\label{fig4}
\end{figure}

When magnetic fields are present, tension forces can restrict these lateral motions. The length and time scales for this restriction depend both on the field strength and geometry.  Studies of magnetized time variable jets with strong cooling show that even initially weak toroidal fields can lead to the collapse of post-shock flows onto the axis \citep{DeColle2008, hansen2015}. Thus, the ring-like structure present at approximately $30 ~pc$ away from the center of the colliding flows can be attributed to the effect of magnetic tension, since it was not seen in our pure hydro case \citep{carroll2014}. Such rings were also seen in other MHD colliding flows runs \citep{vazquez2011, banerjee2009}. We note that in the simulations presented here, the initial field was parallel to the colliding flows.  Thus, as post-shock gas was driven outward from the interaction region, the flow drove arcs in the magnetic field (Fig. \ref{fig4}) whose tension eventually halted further expansion, thereby producing a ring of high density material in the collision plane. The position of this ring can be estimated by assuming the ring has reached a steady-state, as we will do next.

\begin{figure}   
\includegraphics[width=.8\columnwidth]{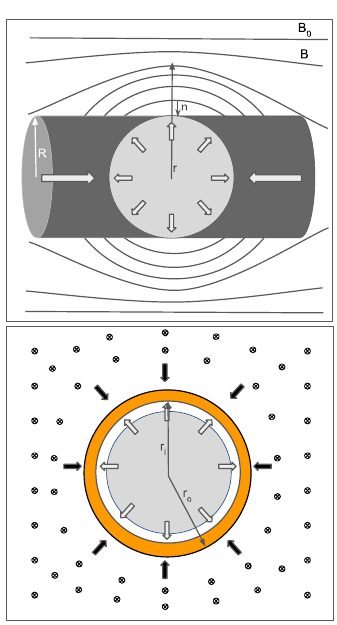}
\caption{Diagram of the magnetic ring model. {\it Top panel} shows the incoming flows, spherical post-shock expansion region and corresponding magnetic field arcs, and the direction along which we are integrating the momentum equation ($r$). Also given is the unperturbed field value ($B_0$), the cylindrical radius of the colliding flows ($R$), and the normal unit vector ($\bold{\hat{n}}$) discussed in the text. {\it Bottom panel} shows the ring of high density material (in the mid-plane of the flows) with inner radius $r_i$ and outer radius $r_o$.}
\label{fig5}
\end{figure}

\subsection{Magnetized Ring Model}

To make calculations simplest, we envision the following scenario. At $t=0$, the magnetic field is $\vec{B}=B_0 \hat{x}$ for $r>R$, where $R$ is the colliding flows radius. For $r<R$, we take $B=0$. This is a fine approximation, given the field is dynamically weak within the flows (recall, $\beta_{ram}\approx38$). As material enters the collision region, it is shocked and then expands away from the collision region, as described above. We approximate this expansion as being spherically symmetric. 

Now, the ram pressure of the ejecta pushes outward on the surrounding low density, magnetized ambient medium. In 2D, this leads to a 'ring' of flux that moves outward (in 3D, a spherical 'shell'; Fig. \ref{fig5}). This ring has two boundaries, an outer radius, $r_o$, and an inner radius, $r_i$ (Fig. \ref{fig5}, {\it bottom panel}). Once this ring comes to steady state, the magnetic pressure at $r_o$ will be balanced by the unperturbed, ambient magnetic pressure, $P_{mag}$, which is $\propto B_0^2$. At $r_i$, $P_{mag}$ will be balanced by the ram pressure of the ejecta. We approximate this ram pressure to be the pre-shock ram pressure, given the low Mach of the flows. Note, we will ignore edge effects near the colliding flows themselves, and instead focus on the dynamics of the ring {\it perpendicular} to the flows (Fig. \ref{fig5}, {\it top panel}).

Assuming the magnetic field is the dominant force in the ambient (therefore, ignoring gravity and pressure forces), the steady-state ideal MHD momentum equation for the ring is given by,

\begin{equation}\label{B_r}
  \nabla \frac{B^2}{2 \mu} = \frac{B^2}{\mu r}\hat{\bold{n}} 
\end{equation}

\noindent where $\mu$ is the magnetic permeability, $r$ is the radius of curvature, which we take to just be the distance from the center of the colliding flows to a position in the ring, and $\bold{\hat{n}}$ is a unit vector normal to the field line that is anti-parallel to $r$ (i.e. $\bold{\hat{n}}=\bold{\hat{-r}}$). Defining the radius of curvature in this way is equivalent to assuming the field is being bent along {\it spherical} arcs. Equation \ref{B_r} says that in steady state, the magnetic tension of the curved field lines (RHS) is balanced by the gradient in magnetic pressure (LHS). Projecting Eqn. \ref{B_r} onto the $r$-axis and integrating gives the following expression for $B^2(r)$ in the ring, 

\begin{equation}
  B^2 = B_0^2(\frac{r_o}{r})^2 
\end{equation}

\noindent Using this, we balance the magnetic and ram pressures at the ring's inner radius $r_i$,

\begin{equation}
  \frac{B_0^2}{2\mu_0}(\frac{r_0}{r_i})^2 = \rho v^2(\frac{R}{r_i})^2 
\end{equation}
 
\noindent which reduces to,

\begin{equation}\label{r_outer}
 {r_0 = \sqrt{\beta_{ram}} R}
\end{equation}

\noindent  Now, the inner radius is where ejected material is being decelerated, so we wish to find an estimate for this edge of the ring to compare with our simulations. For this, we will use flux-freezing. Equating the flux in the ring at steady state,

\begin{equation}
  \theta_1 = \int_{r_i}^{r_o} \frac{B_0 r_0}{r} 2 \pi r dr = 2 \pi B_0 r_o (r_o - r_i)
\end{equation}

\noindent to the flux in the ring initially, 

\begin{equation}
  \theta_2 = \int_{R}^{r_o} B_0 2 \pi r dr = \pi B_0(r_o^2 - R^2)
\end{equation}

\noindent gives,

\begin{equation}
 r_i = \frac{r_o^2 + R^2}{2r_o}
\end{equation}

\noindent Plugging in for $r_o$ (Eqn. \ref{r_outer}) yields,

\begin{equation}
 r_i = R \frac{(\beta_{ram} + 1)}{\sqrt{4 \beta_{ram}}}
\end{equation}

\noindent which is approximately, 

\begin{equation}
 \boxed{r_i \approx \frac{1}{2} \sqrt{\beta_{ram}} R}
\end{equation}

\noindent or $r\approx 3 R$. As seen from the left hand panel of Fig. \ref{fig3}, this simple model reproduces the ring's position to within a factor of two.

\subsection{Main Features in Column Density}

Moving past considerations of the ring, we now focus on the details of the flow in the collision region.  Note that the incoming flows corresponded to WNM, which was at a column density of $N \approx 2\times 10^{20} cm^{-2}$. After passing through the shocks, the gas entered the collision region where the thermal instability drove the gas into the CNM state, which the pdfs of Section \ref{bvn} show was $\approx 500 \times$ denser. We identify the different phases in column density loosely, based on morphology and relative mass fraction from previous studies \citep{heitsch2007, hennebelle2008, banerjee2009}. In addition, we use the minimum $H_I$ column density for  UV-shielding ($N_{HI} \approx 1-2 \times 10^{21} cm^{-2}$) to label regions that have effectively become 'molecular' \citep{vandishoeck1988, vandishoeck1998}. Note that to have accurately tracked molecular gas, we would have needed to include UV shielding in the code. In the images, dark blue regions are the initial WNM, cyan regions are thermally unstable gas, from which the CNM phase and subsequent molecular clouds form (green-yellow), and denser gas (i.e. clumps) within the molecular phase is shown in orange and red. 

The structures seen within the interaction region of the $\theta=0\degr$ case (Fig. \ref{fig3}) were morphologically similar to those of the hydrodynamic version of these runs (\cite{carroll2014}, fig. 2). We note that the degree of heterogeneous or "clumpy" structure appears to be lower in the present magnetized run.  This impression is in line with previous findings for other MHD studies \citep{heitsch2007, hennebelle2008, heitsch2009, chen2014}, which find that fields tend to produce larger, more coherent filamentary structures, compared to the hydro versions of these simulations.  

The first frame of Fig. \ref{fig3}, taken at $t=10 ~Myr$, shows the gas shortly before the first protoclusters formed. By the next frame ($t=20~Myr$), three protoclusters had formed close enough to each other that two of them overlapped and nearly overlapped the third (as can be seen by zooming in on the figure). The location of these sink particles was off-center from the flow axis, indicating they did not form out of a global collapse mode. Instead, they were within local potential minima, associated with regions of high density gas that condensed out of the background turbulent environment and had become gravitationally unstable. By the last time panel of Fig \ref{fig3}, another protocluster had formed off-axis, roughly $25 ~pc$ away. This sink particle was in a large region of high $N$. Given this sink had the fastest average accretion rate of the $\theta=0\degr$ run protoclusters (Fig. \ref{fig2}), this region represented a large and deep potential well. Finally, we note that there was significant widening of the collision region with time, due to the NTS, KH and cooling instabilities.

\begin{figure*}
\includegraphics[width=\textwidth]{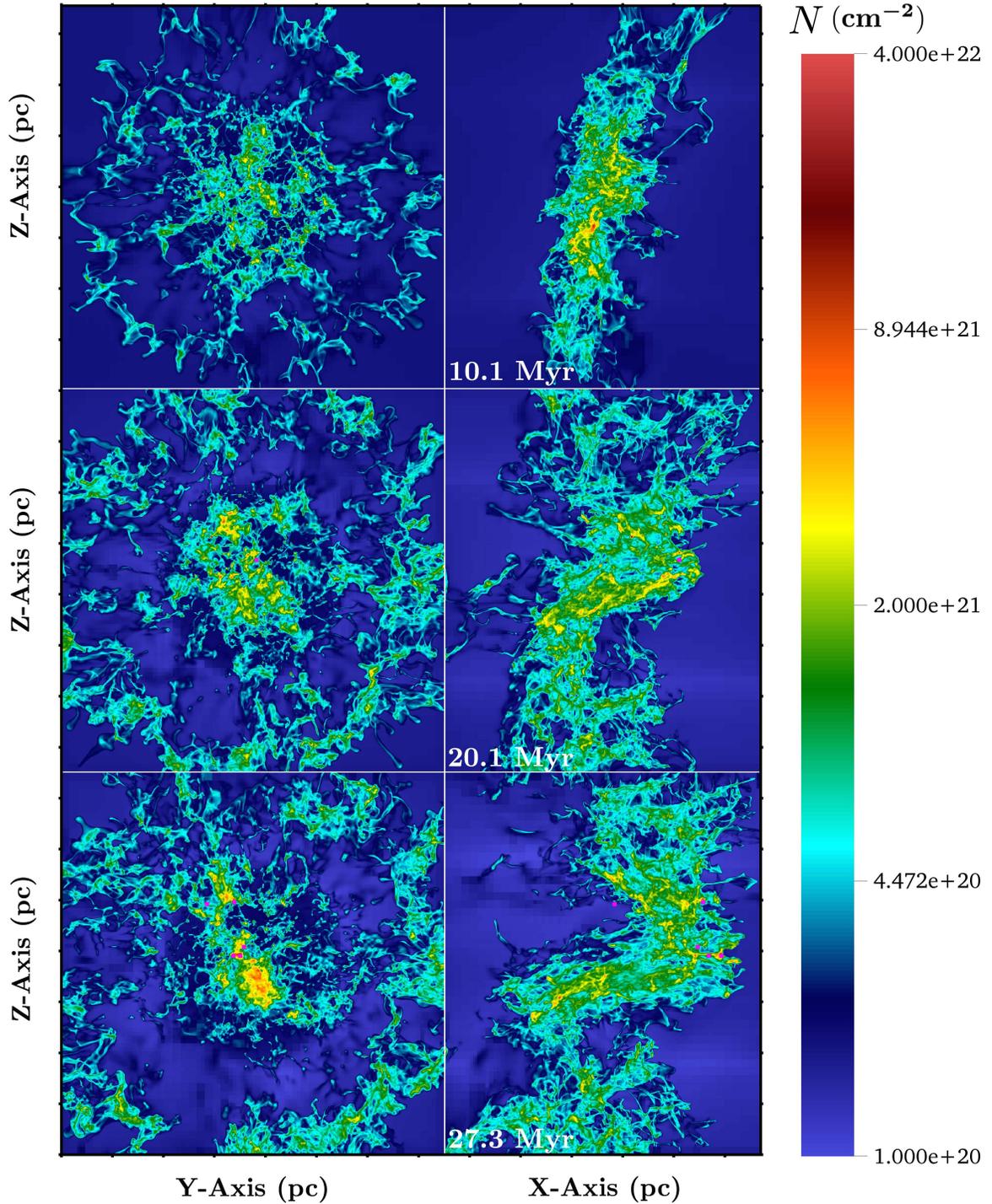}
\caption{Column density map for the $\theta=15\degr$ case. Sink particles are given as fuchsia points. Each tick mark represents $10 ~pc$.}
\label{fig6}
\end{figure*}

We now shift our attention to column density maps of the  $\theta = 15 \degr$ case (Fig. \ref{fig6}). At $t=10~Myr$, the collision region was structurally similar to the $0 \degr$ case, except for regions of lower maximum $N$ (see the left hand column of the plot). Thus, even at low inclination angles, the shear generated by the oblique shocks at the interface disrupted substructure formation to some degree. However, the flows were capable of assembling some localized structures dense enough to become molecular by this time.

By $t=20 ~Myr$, the first protocluster can be seen, having formed at $t_{sink}\approx 17 ~Myr$. It is located near the axis as seen in the left hand panel of Fig. \ref{fig6}, but off-center, in one of the NTSI nodes as seen in the right-hand panel.  Over time, the fingers of the NTSI grew, and by $t=27.3 ~Myr$ the instability had become 'z-shaped', with a large component aligned with the flow axis. Also by this time, four additional protoclusters had formed. Two of these were in the same NTSI node mentioned above, and the other two were roughly $10.2$ and $21.9~ pc$ away, near the edge of the cylinder. The positions of all of these sink particles was again consistent with local instability being triggered in the flow.

\begin{figure*}
\includegraphics[width=.9\textwidth]{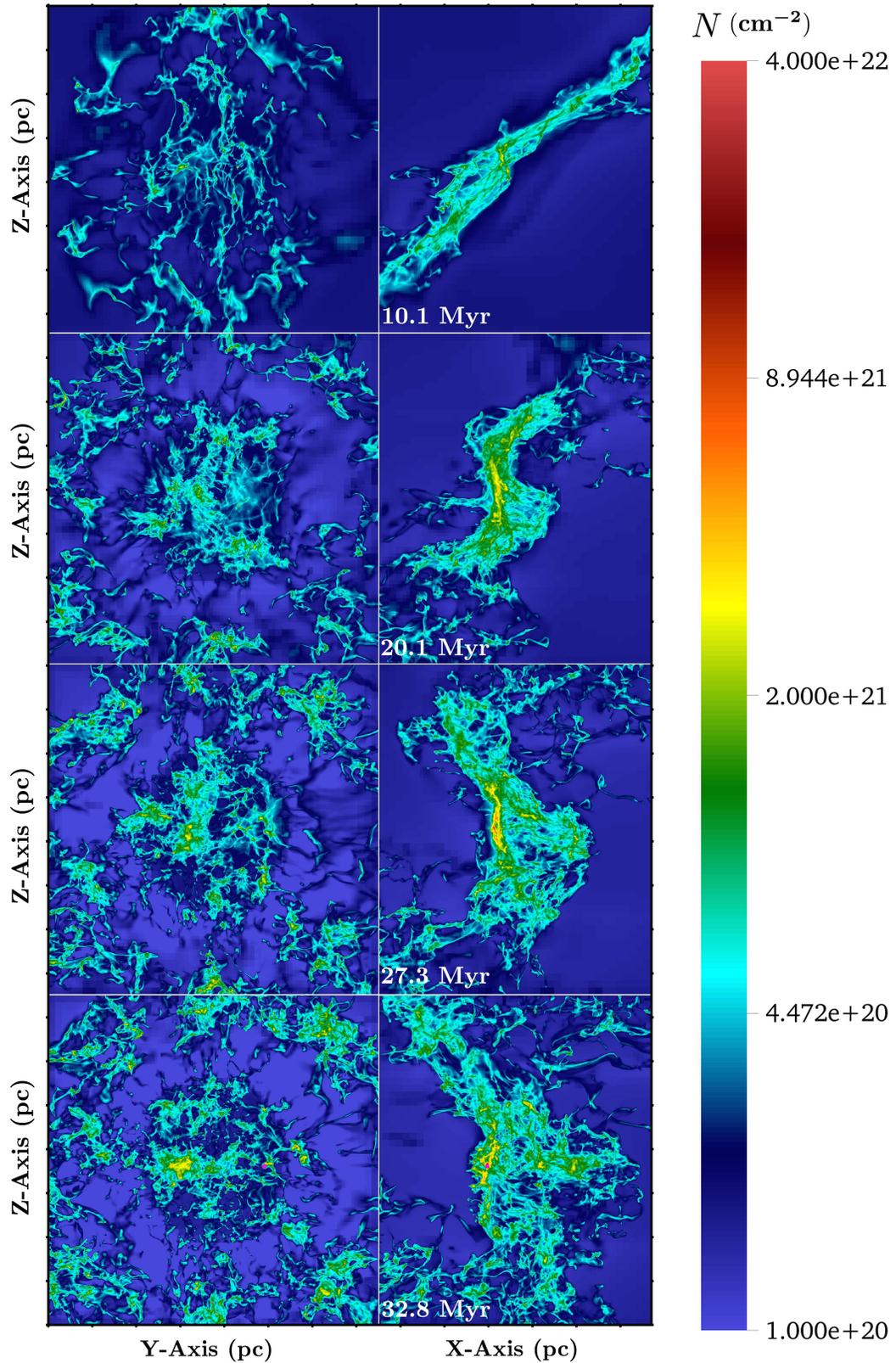}
\caption{Column density map for the $\theta=60\degr$ case. The single sink particle  is given as a fuchsia point. Each tick mark represents $10 ~pc$.}
\label{fig7}
\end{figure*}

Finally, we discuss the extreme $\theta=60\degr$ case, which exhibited very different behavior than the other runs. In particular, we found that the initial steep angle of the collision interface reoriented over time  and became normal to the incoming flows. Animations\footnote{http://www.pas.rochester.edu/~erica/movies.html} of the runs suggested the reorientation was due to perturbations at the collision interface seeding the NTSI. However, field line tension may have also played a role, as a similar, albeit weaker realignment is seen in the hydro version of this run (Haig \& Heitsch in prep).  The reorientation appears to begin in a region where the flow was nearly normal to the surface of an NTSI node  (see the right panel of Fig. \ref{fig7} at $t=10.1 ~Myr$). Because of the high angle of the collision interface, most incoming material was deflected by the oblique shocks to flow parallel along the shock face.  The NTSI node created a local distortion of the shock, allowing oppositely driven material to meet at lower obliquity. The gas behind this stronger shock had higher thermal pressure and began expanding in the (y,z) plane.  This expansion increased the area of the low obliquity regions of the flow, which led to more high pressure post-shock gas as seen by $t=20.1 ~Myr$ (Fig. \ref{fig7}, right).  In this way, it appears the initially highly oblique shock region was transformed into a shock that was more normally directed, relative to the incoming flows. By $t=32.8$, the shock had become nearly fully normal. 

We conclude with comments on the general characteristics of the $\theta = 60 \degr$ collision interface. First, the collision region was less dense than in the other runs at $t=10.1 ~Myr$ (Fig. \ref{fig7}, left). This suppression of growing high density regions arose from the strong shear. Incoming material was sharply deflected away from the axis of the cylinder, and thus less gas accumulated at early times. Second, the collision interface was thinner compared to the other runs at early times, because of this deflection of material away from the interaction region. Material streamed out into the ambient medium, rather than building up along the flow axis. Only by $t=20 ~Myr$, given the reorientation of the shocks bounding the interaction region (and corresponding weaker degree of shear), did material begin to accumulate in the collision region (Fig. \ref{fig7}, left). By $t=32.8 ~Myr$, local collapse had set in and a single protocluster is visible on the CDM, again having formed away from the global potential minimum.

\section{Magnetic Fields and Dynamics: $\bbeta^\mathbf{-1}$ maps}\label{beta}

We next discuss maps of average $\beta^{-1}$, which were generated by discretely summing $\beta^{-1}$ through the grid along the same lines of sight used in the column density maps of the previous section, i.e., 

\begin{equation}
\bar{\beta}^{-1}=\frac{1}{L}\sum_{i=1}^{i=m_x}\beta_i^{-1} dx_i
\label{beta_def}
\end{equation}

 \noindent where $L$, $m_x$, $\beta_i^{-1}$, and $dx_i$ are the length of the box along the given dimension, number of cells along that dimension, value of $\beta^{-1}$ at the $ith$ cell, and the $ith$ cell's width, respectively.
 
\begin{figure*}
\includegraphics[width=\textwidth]{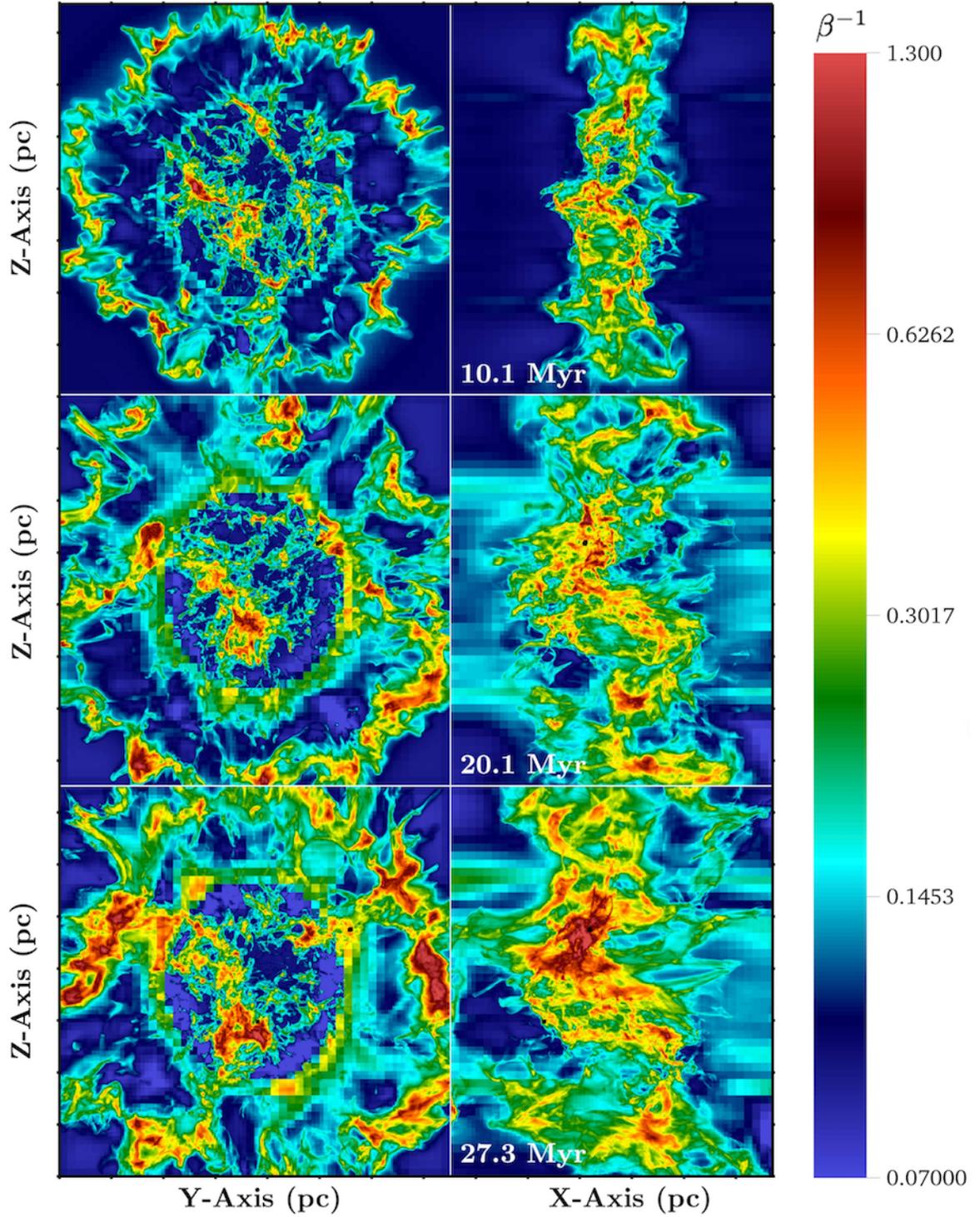}
\caption{Average $\beta^{-1}$ map for the $\theta= 0\degr$ case (see Eqn. \ref{beta_def} for definition). Sink particles are given as black points. Each tick mark represents $10 ~pc$.}
\label{fig8}
\end{figure*}

 Beginning with the $\theta=0\degr$ case, field amplification associated with the radial ejection of gas from the collision region (discussed in Section \ref{morphology}) produced a ring of high average $\beta^{-1}$ at $r\approx 30 ~pc$ (Fig. \ref{fig8}, left). Additionally, there was another, inner ring present from $t=10.1 ~Myr$ on at a distance of $r\approx 15~pc$ from the center of the colliding flows that did not appear in the CDMs. This ring was not contained in the collision region itself, as can be seen from the right-hand frames of Fig. \ref{fig8}. Rather, it stretched down the length of the cylinder, peaking in brightness near the outer edge.  This indicates it was generated by the strong shear present at the boundary between the colliding flows and the stationary ambient medium. This shear layer would have caused strong cooling and therefore a decrease in thermal pressure (hence the increased $\beta^{-1}$ along this boundary).

Looking down the barrel of the flows at $t=10.1 ~Myr$ (Fig. \ref{fig8}, left), there are regions inside of the collision region where average ${\beta}^{-1}$ had increased above its initial value of $\beta^{-1}=0.1$. A glance back at Fig. \ref{fig3} shows these regions are co-located with regions of high column density $N$. In some of these regions, average $\beta^{-1}$ had increased by a factor of $10$ or more.  Since this material is associated with the CNM phase (as discussed in Section \ref{morphology}), its thermal pressure was {\it at least} equal to its initial value (cf. Section \ref{bvn}). Thus, in order for $\beta^{-1}$ to have increased by this amount, the magnetic pressure must have increased by a factor of $10$ or more. This supports that the field was being strongly amplified in the high density gas. 

In addition, we see regions where average $\beta^{-1}$ was reduced below its initial value. These 'voids' (shown in the lowest color on the color bar) generally correspond with regions of low $N$. Since flux freezing implies the accumulation of flux is co-extensive with regions of high density, the voids were formed as material was swept away by both turbulence and the collapse of gas into neighboring potential minima. In other words, voids are associated with regions that have a net positive $\nabla \cdot v$. Thus, regions of low $N$ should not be strongly magnetized, consistent with the voids in Fig. \ref{fig8}. As the simulation evolved, there was an increase in the size of both the magnetic voids, as well as, regions of high average $\beta^{-1}$.

The ramifications of magnetic fields on the dynamics of the flow were already discussed in Section \ref{massform}. The presence of high average $\beta^{-1}$ regions co-located with regions of high $N$ support that magnetic fields suppressed star formation {\it locally}. This is further supported by the location of forming protoclusters. As seen in the $t=20.1$ and $27.3 ~Myr$ panels, protoclusters formed away from regions of highest average $\beta^{-1}$. That is, protocluster formation was inhibited where the field was the strongest. Instead, protoclusters formed where the gas was dense, but average $\beta^{-1}\lesssim0.6$. 

\begin{figure*}
\includegraphics[width=\textwidth]{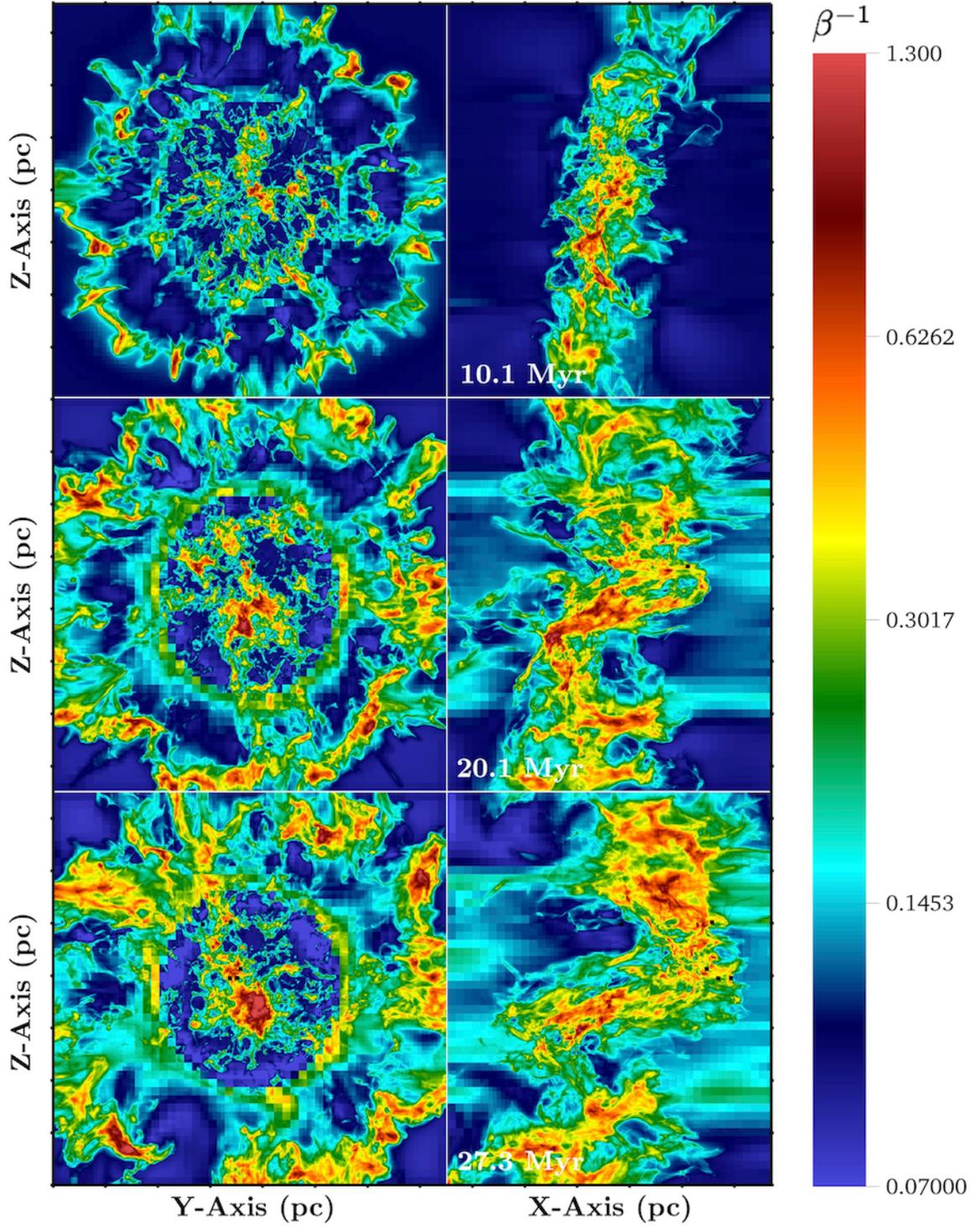}
\caption{Column $\beta^{-1}$ map for the $\theta= 15\degr$ case. Sink particles are given as black points. Each tick mark represents $10 ~pc$.}
\label{fig9}
\end{figure*}

Average $\beta^{-1}$ maps of the $\theta=15\degr$ case (Fig. \ref{fig9}) are similar to the case with no shear. Regions of high average $\beta^{-1}$ are associated with high $N$ structures. There was also a similar formation of magnetic voids. The shear angle present in these runs does, however, leave a signature in the average $\beta ^{-1}$ maps early in the simulation. Looking down the axis of the flows at $t=10.1 ~Myr$, it is evident that there are weaker regions of enhanced average $\beta^{-1}$ (Fig. \ref{fig9}, left). This likely occurs because of the way incoming flows were redirected away from the axis, due to the oblique shocks.  Whereas  material was driven away from the collision region with radial symmetry in the no-shear case, material here picked up positive and negative $v_z$ components across the contact discontinuity. The field lines threading the interaction region must have therefore undergone a local stretching and may have been shorted out by numerical diffusion. This is in contrast to the large scale arcs of field which formed in the purely radial flows of the $\theta =0\degr$ case.

\begin{figure*}
\includegraphics[width=.83\textwidth]{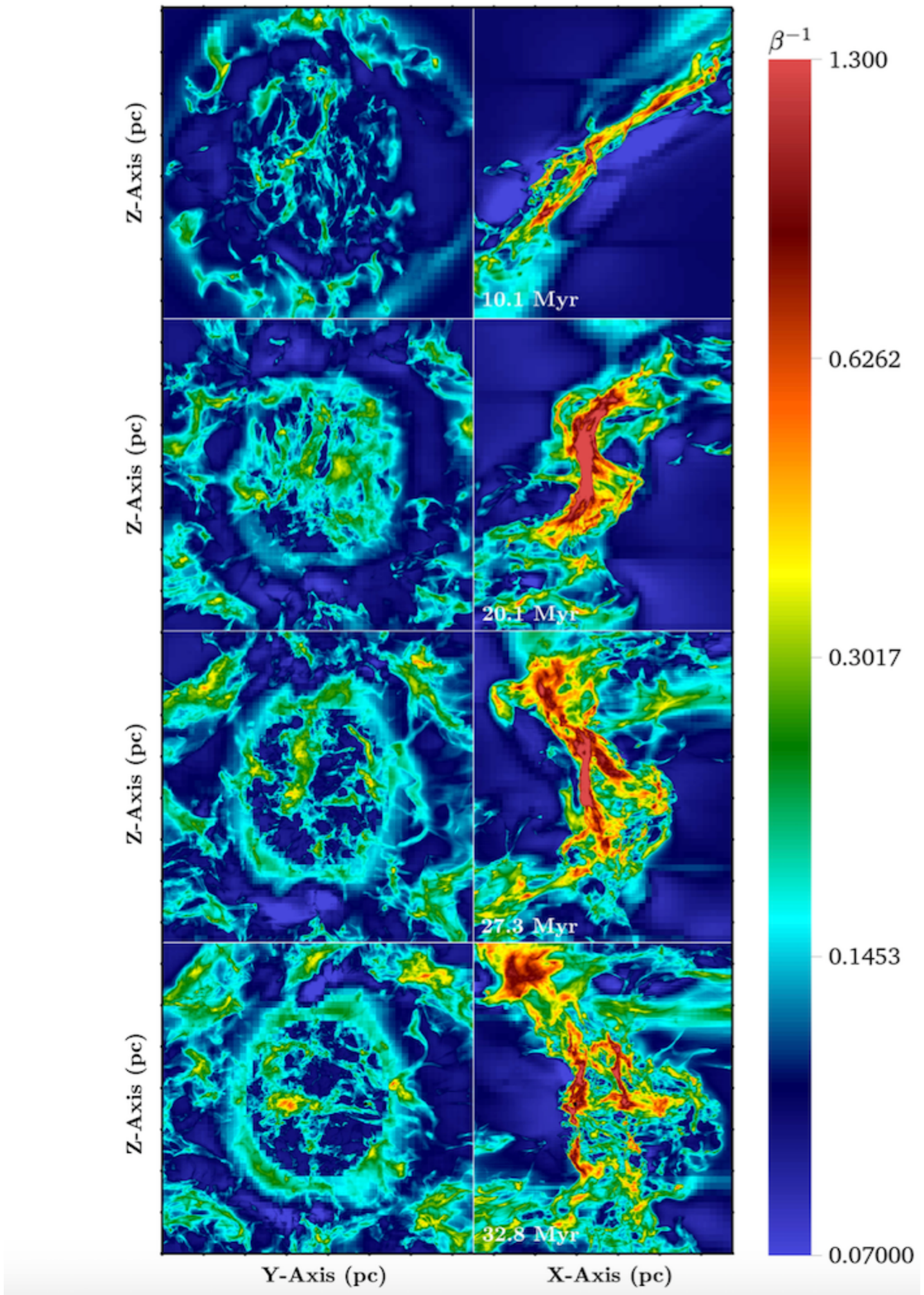}
\caption{Column $\beta^{-1}$ map for the $\theta= 60\degr$ case. The single sink particle  is given as a black point. Each tick mark represents $10 ~pc$.}
\label{fig10}
\end{figure*}

Fig. \ref{fig10} shows a very different set of average $\beta^{-1}$ maps for the $\theta=60\degr$ case. First we note that projected along the axial view we see almost no regions of high average $\beta^{-1}$.   This is because the interaction region remained quite thin early on as gas passed through the oblique shock and was quickly shunted away from the central region.  When seen from the side, however, (Fig. \ref{fig10}, right) we do find significantly higher values of average $\beta^{-1}$ in the collision region. This reflects both the projection through the thin interaction region, as well as the strong local field amplification that occurred due to the shear.  Field lines must have assumed a "z" shaped configuration as gas moving from the left on one side of the contact discontinuity was driven upward by the oblique shock it encountered, and gas moving from the right on the other side of the contact discontinuity was driven downward by the oppositely oriented oblique shock it encountered.

Note that as the collision region reoriented from $t=20.1 ~Myr$ onward  (cf. Section \ref{morphology}), more material and field collected inside of the interaction region in the same manner as was discussed in the lower shear cases. This accounts for the increased regions of high average $\beta^{-1}$ seen in the left hand side of Fig. \ref{fig10} at later times, although overall these regions were much smaller than their lower shear counterparts. 

\section{Thermodynamics - Probability Distribution Functions}\label{bvn}

We now discuss the thermodynamic evolution of the gas as it was first shocked in the flow collision and then cooled into a cold, dense state, which could then undergo further compression (or expansion) due to gravity and/or magnetic fields. The probability distribution functions (pdfs) in this section give the amount of mass at a given pressure and number density. The $y$-axis gives both scaled magnetic {\it and} thermal pressure (marked by the color and gray scale, respectively), and the $x$-axis gives number density $n ~(cm^{-3})$. Isotherms are straight lines of slope $m=1$ on these pdf log-log plots, and increase in temperature from right to left. 

\begin{figure*}
\includegraphics[width=\textwidth]{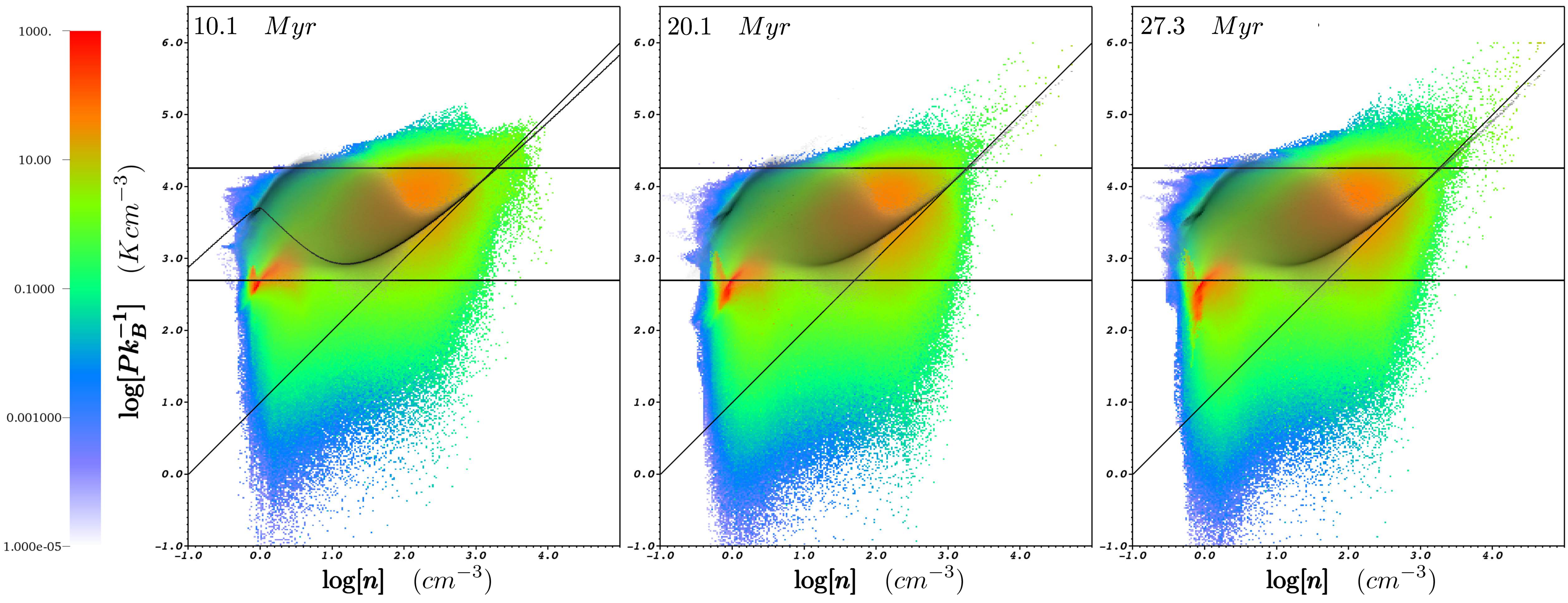}
\caption{Pressure vs. density histograms for the $\theta=0\degr$ case. The y-axis gives both the thermal ({\it grey scale distribution}) and magnetic ({\it color scale distribution}) pressure as a function of number density (the logs of these quantities, that is). The color bar gives the amount of mass at a given pressure and density. The thermodynamic equilibrium  curve is given on the plot, as well as the initial magnetic pressure of the flows ({\it bottom horizontal line}), ram pressure of the flows ({\it top horizontal line}), and $T=10~K$ isotherm ({\it diagonal line}). }
\label{fig11}
\end{figure*}

We begin with the $\theta=0 \degr$ run. The thermal pressure distribution (i.e. {\it grey-scale pdf}, Fig. \ref{fig11}) in this and the remaining runs was identical to previous colliding flows studies. Namely, gas was shocked and heated as it entered the collision region and via the thermal instability, cooled down until it reached the equilibrium curve. For a more detailed discussion of the dynamics relayed by the thermal pressure pdf, we refer the reader to previous work (e.g. \cite{carroll2014}). Presently, we consider the simultaneous thermal and magnetic evolution of the gas as shown by overlaying the $P_{mag}$, $n$ pdf in color. Note, however, it is not always possible to {\it exactly} correlate values in one pdf with values in the other (i.e. to simultaneously know the $P_{therm}$ of a given $P_{mag}, n$ combination). 

As can be seen in Fig. \ref{fig11}, {\it most} of the mass at low densities (i.e. near $\log n = 0$) had $\beta > 1$, consistent with the notion of magnetic voids discussed in Section \ref{beta}. This follows from comparing the $P_{mag}$ distribution with the $P_{therm}$ distribution at these densities. While the distribution of $P_{mag}$ had a much higher spread, this spread was {\it mostly} below the $P_{therm}$ distribution. In contrast, gas at higher densities ($\log n > 2.5$) was mostly on top of the equilibrium curve, implying most of the densest gas had $\beta <1$. This is in agreement with the densest gas having had enhanced magnetic support (and thus greater support against collapse), as was also discussed in Section \ref{beta}. Despite the spread in $P_{mag}$ at these high $n$, the gas was entirely at $T=10 K$, as the $P_{therm}$ distribution was constrained to lie on the equilibrium curve at these densities. Lastly, the $P_{mag}$, $n$ pdf tended to shift toward the upper right of the plot while becoming more tightly distributed. If a line were drawn through the middle of the entire distribution, it would be roughly linear until about $\log n \approx 1$, and would then shift upwards, which on this log-log plot would correspond to a power law relationship similar to what is observed in other work \citep{heitsch2007, hennebelle2008, banerjee2009}. 

Now we discuss the dynamics attributable to these various features of  the $P_{mag}$, $n$ pdf. As the simulation proceeded, turbulence at the collision interface (produced by the KH and NTS instabilities) generated magnetic field fluctuations. Reconnection diffusion, effective at removing excess magnetic energy from magnetized, turbulent flows (\cite{lazarian2010}; see also \cite{klessen2000, federrath2011}), created the {\it large} (i.e.  many orders of magnitude) spread in $P_{mag}$ at low $n$. This spread in $P_{mag}$ persisted as the gas was cooled and compressed. Flux-freezing, which accompanied this compression, then led to a sharp increase in $P_{mag}$, as seen by the rather large island of material (in red shading) up and to the right of the initial values. Note that colors correspond to a log scale, and so the the amount of mass within this population was significantly greater than any other $P_{mag}, n$ combination of the flow. 

As time continued, a high $P_{mag},n$ ''tail'' of the pdf formed that, 1) followed the equilibrium curve, 2) was above the ram pressure line of the flows, and 3) only appeared after protoclusters had formed in the flow. Thus, this tail traced gravitational instability in the flow. Most of the mass in the tail was at a $\beta < 1$, consistent with the discussion from Section \ref{beta} that regions of high density were correlated with regions of high $\beta^{-1}$. Additionally, there was a spread in $\beta$ for the population of gas parcels in this tail, as was also discussed in Section \ref{beta}. In particular, protoclusters formed in dense {\it enough} gas that was at lower average $\beta^{-1}$ than the surroundings. Numerical reconnection within this high $P_{mag}$, $n$ tail reduced the local field strength, thus enabling protocluster formation. 

\begin{figure*}
\includegraphics[width=\textwidth]{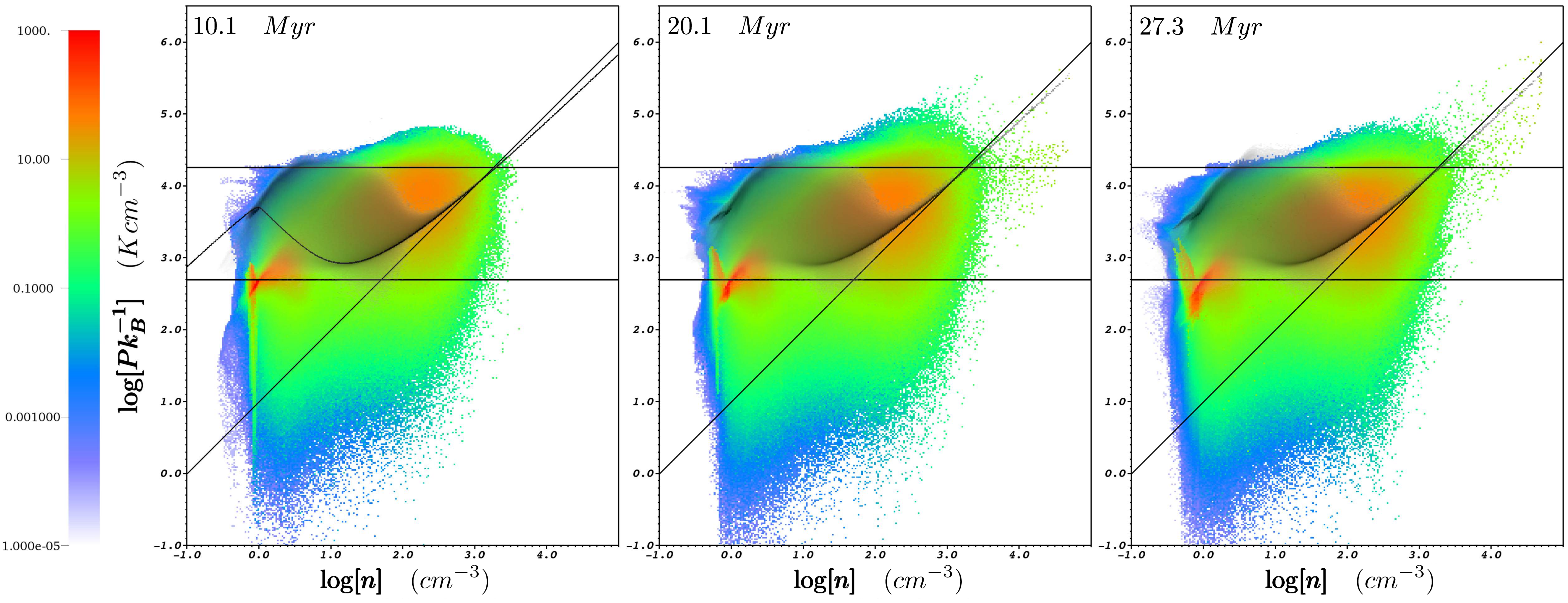}
\caption{Pressure vs. density histograms for the $\theta=15\degr$ case. As before, the y-axis gives both the thermal ({\it grey scale distribution}) and magnetic ({\it color scale distribution}) pressure as a function of number density (the logs of these quantities, that is). The color bar gives the amount of mass at a given pressure and density. The thermodynamic equilibrium  curve is given on the plot, as well as the initial magnetic pressure of the flows ({\it bottom horizontal line}), ram pressure of the flows ({\it top horizontal line}), and $T=10~K$ isotherm ({\it diagonal line}). }
\label{fig12}
\end{figure*}

The pdfs for the $\theta=15 \degr$ case (Fig. \ref{fig12}) were similar to those for the $\theta=0\degr$ case. The largest difference was the delayed growth of the high $P_{mag},n$ tail in the higher shear run. Thus, the plot offers further evidence that shear suppressed compression in the flows. Eventually collapse was triggered in localized pockets of the flow, allowing  gas parcels to begin to climb the equilibrium curve, as seen by $t=20.1 ~Myr$.  As in the previous case, this coincided with the presence of newly formed protoclusters in the flow. 

\begin{figure*}
\includegraphics[width=\textwidth]{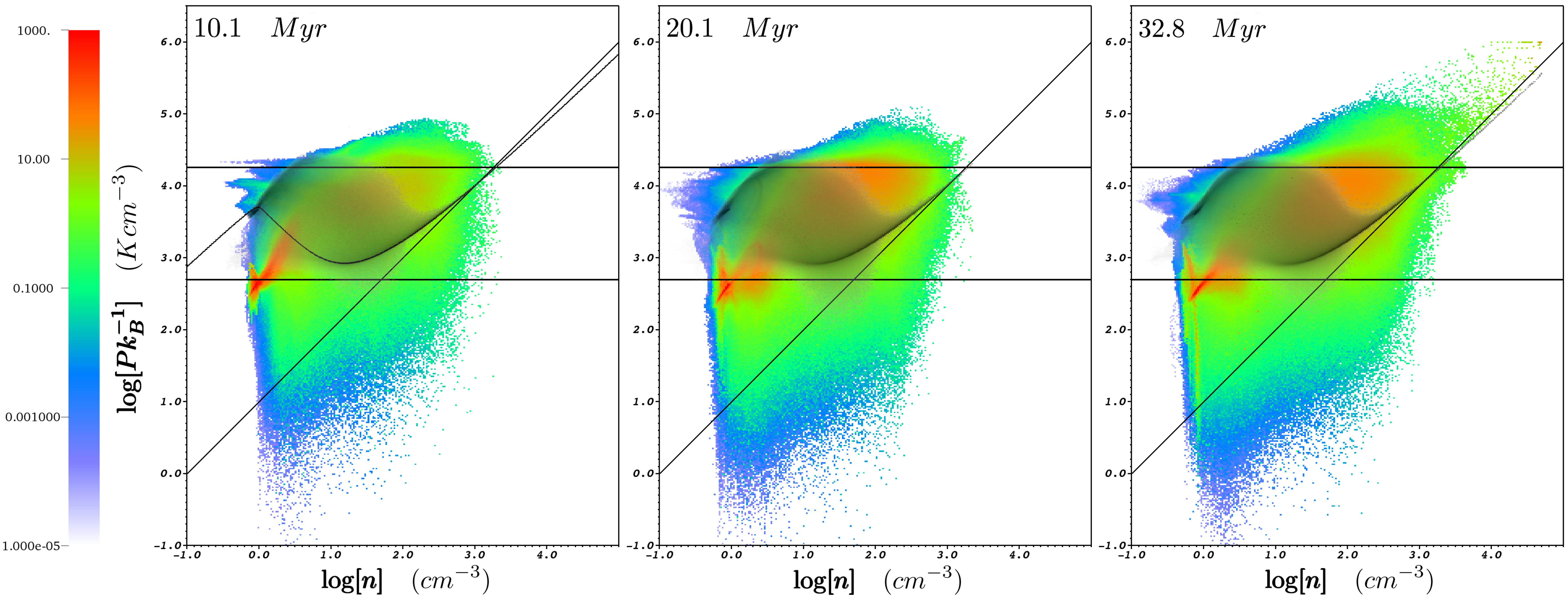}
\caption{Pressure vs. density histograms for the $\theta=60\degr$ case. As before, the y-axis gives both the thermal ({\it grey scale distribution}) and magnetic ({\it color scale distribution}) pressure as a function of number density (the logs of these quantities, that is). The color bar gives the amount of mass at a given pressure and density. The thermodynamic equilibrium  curve is given on the plot, as well as the initial magnetic pressure of the flows ({\it bottom horizontal line}), ram pressure of the flows ({\it top horizontal line}), and $T=10~K$ isotherm ({\it diagonal line}). Note, the pdfs were similar enough between $t=20.1$ and $27.3 ~Myr$ that we do not include the $t=27.3 ~Myr$ panel in the figure. }
\label{fig13}
\end{figure*}

Keeping in mind the trends discussed previously, Fig. \ref{fig13} shows an even longer delay in collapse/compression in the highest shear angle case, $\theta=60 \degr$. This is evident by the diminished upper-right island of high $P_{mag}, n$ material early on. It is also shown by the greater delay in the high $P_{mag}, n$ tail compared to the lower shear runs, which was still not visible by $t=27.3 ~Myr$. Only by $t=32.8 ~Myr$ did the high $P_{mag},n$ tail emerge in the pdf, again corresponding with the presence of a newly formed protocluster in the flow. 

\section{Energy Spectra}\label{spectra}

We next turn to power spectra of the gas in the collision region for kinetic, gravitational, and magnetic energies ($E_{kin}, E_{grav},$ and $E_{mag}$, respectively), following \cite{carroll2014}. To accommodate steeper shear angles, the $x$ dimension of the analysis region increased with $\theta$ and a $10 ~pc$ buffer was added to either side. The dimensions of the analysis region for the different runs are given in Table \ref{tab2}.

\begin{table}
\centering
\caption{Analysis region for the different runs. Each were centered on the collision interface and had lengths in $x$, $y$, and $z$, given by $L_x$, $L_y$, $L_z$. The maximum wavelength $\lambda_{max}$ ($\propto k_{min}^{-1}$) for each run is also given, which was equal to the longest dimension of the analysis region.}
\label{tab2}
\begin{tabular}{@{}clclclc@{}}
\hline
$\theta (\degr)$ & & $L_x ~(pc)$  & & $L_y, ~L_z ~(pc)$ & & $\lambda_{max}$ (pc) \\
\hline
 0 & & 20 & & 40 & & 40   \\
 15 &  & 31  & & 40 & &  40 \\
60 & & 90 & & 40 & & 90  \\
\hline
\end{tabular}
\end{table}

\begin{figure*}
\includegraphics[width=\textwidth]{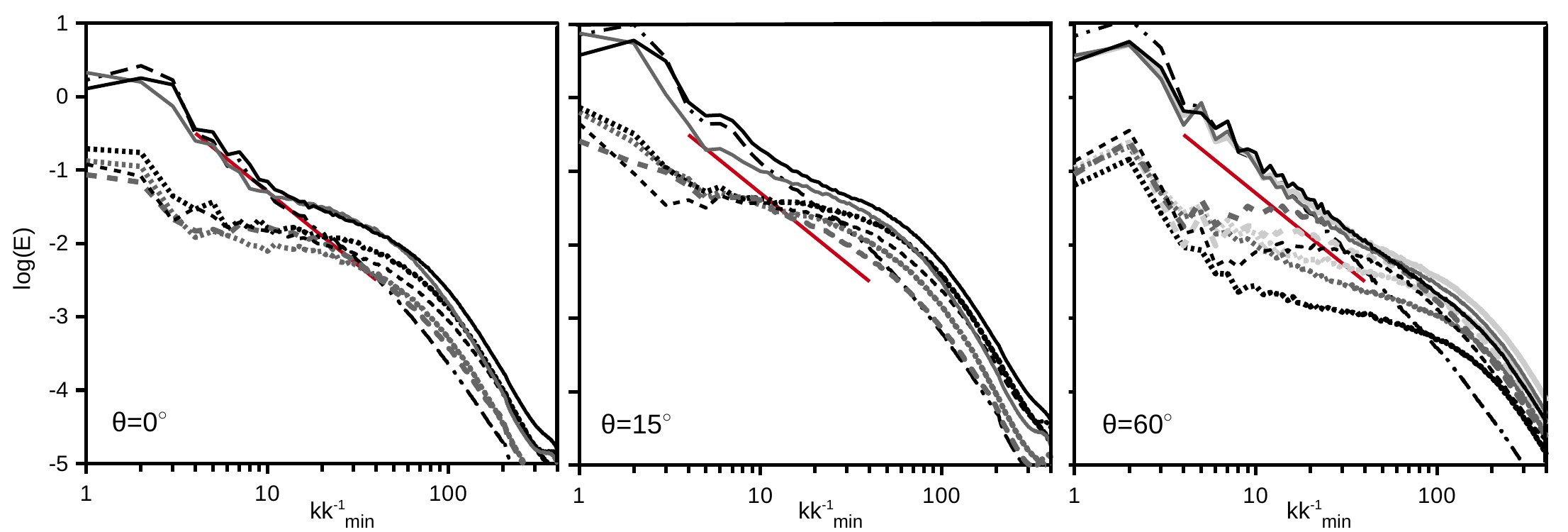}
\caption{Energy spectra for the various runs. Shown are the kinetic ({\it smooth lines}), magnetic ({\it dashed}), and gravitational energies ({\it fine-dashed}) over time (increasing in time from dark to light). There is also a $v^2$ line ({\it dotted-dashed line}) and corresponding $k^{-2}$ line ({\it red line}) to measure the degree of Burger's type turbulence in the runs. For the $\theta=0-15\degr$ plots, black and gray lines are $t=10.1$ and $27.3~Myr$, respectively. For the $\theta=60\degr$ run, black, gray, and light gray lines are for $t=10.1,$ $20.1$, and $32.8~Myr$, respectively.} 
\label{fig14}
\end{figure*}

Spectra for the different runs are shown in Fig. \ref{fig14}, for three different times. Note, the ${x}$-axes in the spectra are equivalent to $\lambda_{max} \lambda^{-1}$, where $\lambda_{max}$ is given in Table \ref{tab2}. Thus, each unit on the $x$-axis represents that fraction of $\lambda_{max}$. We include a $k^{-2}$ line (solid line, red) to compare to the spectra of $v^2$ (dotted-dashed line, black) for $t=10.1 ~Myr$. We do not include the $v^2$ spectrum for other times, as it was largely unchanged throughout the course of the simulations.

As can be seen by the $v^2$ spectrum, gas displayed the Burger's turbulent spectrum ($v^2 \propto k^{-2}$) in all of the runs over some range of $k$. For the $\theta=0\degr$ case, this was $3<k k_{min}^{-1}<40$, corresponding to length scales of roughly $13~pc>\lambda>1~pc$. At the low $k$ end of the spectra, the driving scale of the turbulence is apparent and was on the order of the colliding flows radius ($\lambda\approx 40 ~pc$). At higher $k$, both dissipation and gravitational collapse limited the inertial range captured in the grid.

The dominant energy on all size scales for all of the runs was $E_{kin}$ (smooth, black lines), and remained so throughout the course of the simulations. This was due to the power being generated in the colliding flows themselves. We note that in the hydro version of the $\theta=0\degr$ case \citep{carroll2014}, $E_{kin}>E_{grav}$ at late times in the flow on large scales ($\lambda>10~pc$). This indicated a lack of global collapse. The much stronger suppression of $E_{grav}$ (fine-dashed lines) in the MHD runs (i.e. $E_{kin}>>E_{grav}$) shows that large-scale global collapse was also not occurring in the MHD cases. Also, $E_{kin}<E_{grav}$ on small scales ($\lambda<10 pc$) in the hydro case. This indicated strong local collapse was occurring. That $E_{kin}>E_{grav}$ on small scales in the MHD cases further supports that magnetic fields have impaired local collapse. A diminished degree of local collapse occurred in the MHD runs, and only in regions where $E_{grav}>E_{mag}$.

Indeed throughout the runs, $E_{grav}$ was comparable to $E_{mag}$, on most scales. However, in the $\theta=0\degr$ case, $E_{grav}$ was generally higher than $E_{mag}$ for $10 <k k_{min}^{-1}< 100$. Note, this corresponds to a length scale of $.4<\lambda ~(pc^{-1})<4$, which is the size scale of protoclusters. This is consistent with weak, localized collapse, due to a stronger gravitational field on these scales. By $t=27.3 ~Myr$, there was a slight increase in $E_{mag}$ relative to $E_{grav}$, which may reflect an enhancement of the field due to compression from gravitational collapse. 

The $\theta=15 \degr$ spectra show similar behavior to the $\theta=0 \degr$ case.  At intermediate scales, $E_{mag}$ and $E_{grav}$ were comparable. At smaller scales, as gravitational collapse set-in, $E_{grav}>E_{mag}$.  The effect of shear in this case did not make itself readily apparent in the energy spectra.

The spectra for the $\theta=60 \degr$ case, however, show clearly the effects of imposed shear on the distribution of energies at different scales.  At $10 ~Myr$, $E_{mag}>E_{grav}$, on all size scales. In the range of $9<k k_{min}^{-1}<90$, corresponding now to size scales of $1<\lambda(pc^{-1})<10$ (cf. Table \ref{tab2}), we see $E_{mag} >> E_{grav}$, with $E_{mag}$ approaching $E_{kin}$ near $kk_{min}^{-1} \sim 60$ ($\lambda \sim 1.5 ~pc$). Note that this is the only run in which such an equipartition occurred. This enhanced $E_{mag}$ relative to $E_{kin}$ occurred with a simultaneous decrease in power of $E_{grav}$ on all scales compared to the other runs. Thus, the signature of high shear was a relative amplification of the field on protocluster scales, as well as a decrease in $E_{grav}$ on all scales, which we attribute to the disruption of dense structures forming in the flow as already discussed. Only by the end of the simulation ($t=32.8 ~Myr$) did $E_{grav}$ become comparable with the energy locked up in the magnetic field (which was itself comparable to $E_{kin}$).  Thus, the spectra are consistent with the formation of a sink particle at late times.

\section{Discussion}\label{discussion}

We have presented 3D, adaptive mesh refinement simulations of magnetized, shear colliding flows including self-gravity, sink particles, and cooling. Feedback was not included in the present work and thus our simulations only followed the formation and early evolution of molecular clouds. The simulations were of two colliding flows, intersecting at an inclined interface under a dynamically weak magnetic field. The inclined interface was used to generate shear at the collision layer, and was varied from not-inclined (normal incidence of the colliding flows) to highly inclined. As molecular clouds are unlikely to form from perfectly head-on collisions between two large-scale streams of gas, breaking the symmetry of the collision interface in this way was a previously unexplored next step in the colliding flows model exploration. The purpose of this set of experiments was to study the effect of shear on molecular cloud formation in magnetized colliding flows.

Our work has shown that shear, imposed by an inclined collision interface, impacts cloud dynamics and reduces protocluster formation. In particular, as shear increased (i.e. the inclination angle steepened), {\it it took longer to form protoclusters, they formed in lower numbers, and they were lower in mass due to diminished accretion rates}. These effects are consistent with recent work by \cite{kortgen2015}, who found that higher degrees of inclination between colliding flows leads to a reduced number of sink particles, as well as a delay in their formation. Additionally, \cite{chen2014} show that under ideal MHD conditions, increasing the angle between upstream colliding flows and the magnetic field (which, with a change of reference frame is similar to our setup here) decreases the number of gravitationally bound cores that form in the post-shock region. 

We also showed that without shear, protocluster formation was greatly impeded in the presence of even a weak field ($\beta=10,~ \beta_{ram}\approx 38$) alone. This was evidenced by the stark difference in protocluster number between the no-shear, MHD case ($\theta=0\degr$) and the hydro version of this simulation in \cite{carroll2014}. More than $5$ times as many protoclusters formed in the hydro run (n=27) compared to the MHD run (n=4). This result fits in with other colliding flows studies that have shown magnetic fields impair gravitational collapse. \cite{hennebelle2008} showed that it takes longer to form self-gravitating objects in MHD colliding flows (aligned field, $B_0=10 ~\mu G$), compared to flows without a field. \cite{heitsch2009} also compared colliding flows with and without a magnetic field and found that the degree of post-shock turbulence decreases with a magnetic field. This led to a suppression of dense cores forming in the magnetized cases. Additionally, \cite{vazquez2011} found that increasing the magnetic field strength in colliding flows simulations decreases the SFE of forming clouds.

To understand the flow dynamics responsible for the weaker protocluster formation, we began by distinguishing between 'global' and 'local' collapse, and showed that collapse in our simulations was due to {\it localized} regions becoming unstable, rather than the entire collision region. Indicators of local collapse included the positions of sink particles away from the global potential minimum, weak mass 'fall-back' into the collision region, and power spectra that showed $E_{kin}>E_{grav}$ on large scales, but $E_{kin}<E_{grav}$ on  small scales \citep{carroll2014}. Our results were consistent with these indicators: sinks formed away from the center of the collision region, they had variable accretion rates, and both $E_{grav}$ and $E_{mag}$ were $<<E_{kin}$ on large scales, but $E_{grav}\sim E_{kin}>E_{mag}$ on small scales. All of the runs ($\theta=0-60\degr$), as well as the hydro run, exhibited local collapse only (i.e. did not collapse globally). 

Compared to the hydro run, collapse was weakened in our $\theta=0\degr$ case due to the interplay between the turbulent post-shock velocity field and the magnetic field. Turbulence in the interaction region introduced distortions to the magnetic field, which produced regions of local field amplification. These were shown to be co-located with regions of high density (Sections \ref{beta} and \ref{bvn}). Thus, local collapse had an additional form of support in the MHD case(s) -- magnetic support, due to field amplification (see also \cite{banerjee2009, heitsch2009}). Over time, numerical reconnection would have reduced the field strength in these pockets of dense gas \citep{lazarian2010, vazquez2011, federrath2011, chen2014}, thereby allowing highly localized regions to become magnetically supercritical and produce protoclusters (should the gas also be Jeans unstable). 

As shear increased in the flows, a higher degree of post-shock turbulent velocity resulted. This disrupted the formation of dense structures in the flow, which further inhibited local collapse and protocluster formation (by locally decreasing the M2FR, as well as, increasing the Jeans length). This seemed to be the dominant effect of shear, as we did not find that stronger shear led to greater field amplification. The reasons for this were likely two-fold. While the shear generated by the oblique shocks at the collision interface would have led to greater distortions of the field \citep{hartmann2001, heitsch2007, chen2014}, they were also likely shorted out faster in the higher shear cases. Additionally, given the disruption of high density structures forming in the flow, regions of increased field strength due to flux freezing would have also been diminished. Both the pdfs and spectra (Sections \ref{bvn} and \ref{spectra}, respectively) support the conclusion that field amplification did not increase with shear. They do, however, show decreases in high density structures. This is consistent with \cite{kortgen2015}, who attribute declining sink particle formation to shear disrupting the formation of high density structures. 

Lastly, we discussed the large-scale realignment of the collision interface that took place in our highest shear angle case, $\theta=60 \degr$. This unexpected result may have arisen from the orientation of an NTSI node with respect to the oncoming flows, as discussed in Section \ref{morphology}. However, tension in the magnetic field may have also contributed to the realignment. This is supported by current work by Haig \& Heitsch (in prep), which shows that the steep $60 \degr$ inclination angle reorients to a {\it lesser} degree {\it without} a magnetic field. A realignment of the collision interface was also mentioned in \cite{kortgen2015}.

\section*{Acknowledgments}

We thank Ralph Klesson, Simon Glover, Mordecai-Mark Mac Low, Eric Mamajek, Alice Quillen, and Andy Pon for interesting discussions on the present work. 
We thank Christoph Federrath and Clare Dobbs for their valuable comments and suggestions, and Bastian K\"{o}rtgen for proposing including an early mass distribution of protoclusters. We thank the referee, Robi Banerjee, for his insightful review and recommending a vorticity analysis. We thank the University of Rochester's Center for Integrated Research Computing (CIRC) for time on their supercomputers, data storage, and visualization at the VISTA collaboratory, and the Texas Advanced Computing Center and XSEDE for computing time. Lastly, we thank Rich Sarkis of the University of Rochester for maintaining our data systems. This work was supported by the U.S. Department of Energy through grant GR523126, the National Science Foundation through grant GR506177, and the Space Telescope Science Institute through grant GR528562. 

\bibliography{erica}

\newpage
\bsp

\label{lastpage}

\end{document}